\documentclass[preprint,12pt,sort&compress]{elsarticle}

\usepackage{amsmath,amssymb}
\usepackage[]{graphicx}
\usepackage[latin1]{inputenc}
\usepackage{color}
\usepackage{enumerate}
\usepackage{hyperref}
\hypersetup{colorlinks=true,citecolor=blue}

\journal{Chemical Engineering Science}

\begin{document}

\begin{frontmatter}

\title{Estimation of Particle Size Distribution and Aspect Ratio of Non-Spherical Particles From Chord Length Distribution}

\author[a1]{Okpeafoh S. Agimelen\corref{cor1}}
\ead{okpeafoh.agimelen@strath.ac.uk}

\author[a2]{Peter Hamilton}

\author[a3]{Ian Haley} 

\author[a2]{Alison Nordon}

\author[a4]{Massimiliano Vasile}

\author[a1]{Jan Sefcik}
                                                                                            
\author[a5]{Anthony J. Mulholland}
\ead{anthony.mulholland@strath.ac.uk} 

\cortext[cor1]{Corresponding author}

\address[a1]{Department of Chemical and Process Engineering, University of Strathclyde, James Weir Building, 75 Montrose Street, Glasgow, G1 1XJ, United Kingdom.}

\address[a2]{WestCHEM, Department of Pure and Applied Chemistry and Centre for Process Analytics and Control Technology, University of Strathclyde, 295 Cathedral Street, Glasgow, G1 1XL, United Kingdom}

\address[a3]{Mettler-Toledo Ltd., 64 Boston Road, Beaumont Leys Leicester, LE4 1AW, United Kingdom}

\address[a4]{Department of Mechanical and Aerospace Engineering, University of Strathclyde,  James Weir Building, 75 Montrose Street, Glasgow, G1 1XJ, United Kingdom.
}

\address[a5]{Department of Mathematics and Statistics, University of Strathclyde, Livingstone Tower, 26 Richmond Street, Glasgow G1 1XH, United Kingdom
}

\begin{abstract}
Information about size and shape of particles produced in various manufacturing processes is very important for process and product development because design of downstream processes as well as final product properties strongly depend on these geometrical particle attributes.  However, recovery of particle size and shape information in situ during crystallisation processes has been a major challenge.  The focused beam reflectance measurement (FBRM) provides the chord length distribution (CLD) of a population of particles in a suspension flowing close to the sensor window. Recovery of size and shape information from the CLD requires a model relating particle size and shape to its CLD as well as solving the corresponding inverse problem. 

This paper presents a comprehensive algorithm which produces estimates of particle size distribution and particle aspect ratio from measured CLD data. While the algorithm searches for a global best solution to the inverse problem without requiring further a priori information on the range of particle sizes present in the population or aspect ratio of particles, suitable regularisation techniques based on relevant additional information can be implemented as required to obtain physically reasonable size distributions. We used the algorithm to analyse CLD data for samples of needle-like crystalline particles of various lengths using two previously published CLD models for ellipsoids and for thin cylinders to estimate particle size distribution and shape. We found that the thin cylinder model yielded significantly better agreement with experimental data, while estimated particle size distributions and aspect ratios were in good agreement with those obtained from imaging.
\end{abstract}


\end{frontmatter}

\section{Introduction}
\label{intro}

Manufacturing of particulate products in pharmaceutical and fine chemicals industries includes various particle formation processes, such as crystallisation or granulation, and downstream processing of resulting suspensions or powders as well as final product properties are strongly dependent on geometrical particle attributes, most importantly size and shape. Design and operation of particle formation processes greatly benefits from in situ monitoring of particle size and shape, but it has been a major challenge to get reliable quantitative estimates of these key particle attributes, especially in cases where solid loadings are relatively high or sampling is challenging. There are numerous particle sizing techniques, including sieving, electrical zone sensing, laser diffraction, focused beam reflectance measurement (FBRM) and imaging \cite{Washington1992,Heinrich2012}. While several techniques are well suited for determination of particle size distributions of spherical particles, there are significant challenges when particles become strongly non-isometric, such as in the case of needle-like or plate-like particles, which are ubiquitous in pharmaceutical manufacturing. Imaging is well suited for dealing with high aspect ratio particles, but accurate determination of particle size and shape by imaging typically requires highly diluted samples and/or specially designed flow cells, which make it difficult to apply in situ under process conditions. Although laser diffraction and reflectance techniques provide information which is sensitive to particle shape, extracting accurate shape information has been challenging since appropriate models need to be used and corresponding inverse problems need to be solved. Suspensions also need to be relatively dilute for laser diffraction measurements in order to avoid multiple scattering effects.

Reflectance techniques, such as FBRM, are particularly suitable for in situ monitoring of particles in suspensions during the manufacturing process. FBRM measures chord length distribution (CLD), which depends on both size and shape of particles present in a suspension. There has been considerable efforts \cite{Tadayyon1998,Ruf2000,Heath2002,Wynn2003,Worlitschek2003,Worlitschek2005,Li2005n1,Li2005n2,Vaccaro2006,Kail2007,Kail2008,Kail2009,Scheler2013} devoted towards obtaining useful information about particle geometrical attributes from this technique, leading to the development of suitable models \cite{Tadayyon1998,Ruf2000,Heath2002,Wynn2003,Worlitschek2003,Worlitschek2005,Li2005n1,Vaccaro2006,Kail2007,Kail2008,Kail2009,Scheler2013} for CLDs for particles of various shapes in order to obtain particle size distributions from FBRM data. 

However, the inverse problem of retrieving size and shape information from FBRM data is non-trivial \cite{Heinrich2012}. The inverse problem is well-known to be ill-posed, i.e., there are potentially multiple solutions in terms of particle size distributions and shape which give essentially the same CLD within the accuracy of experimental data. Several regularisation approaches have been proposed to deal with this problem \cite{Worlitschek2005,Li2005n1} but there is still a challenge of finding a global best solutions for physically reasonable combinations of particle size distribution and shape. One important factor which can be used to constrain inverse problem solutions is the size range of particles used in the calculations. In the work by Ruf et al \cite{Ruf2000} information about particle size range was obtained by a laser diffraction technique and microscopy, while Worlitschek et al. \cite{Worlitschek2005}, Li et al. \cite{Li2005n2}, Li et al. \cite{Li2013,Li2014} and Yu et al. \cite{Yu2008} obtained particle size range information by sieving. Also, Kail et al. \cite{Kail2009} obtained information about particle size range in their population of particles from the manufacturer. However, information about particle size range may not be readily available or it may not be convenient to obtain this information a priori (for example in a manufacturing process). 

When moving from modelling of CLD of single particles to a population of particles of various sizes, it is necessary to properly account for size effects. It has been previously shown \cite{Simmons1999,Hobbel1991,Vaccaro2006} (see also section 3 of the supplementary information) that probability of larger particles to be detected by the FBRM probe is proportional to their characteristic size. While this effect has been taken into account in some cases \cite{Ruf2000,Worlitschek2005} it has been neglected in some other cases \cite{Li2005n1,Li2005n2} in the previous literature, which may introduce significant errors if the size range of particles in the population is relatively large. 

Early CLD geometrical models \cite{Hobbel1991,Simmons1999,Langston2001,Hukkanen2003,Barrett1999} were based on populations of spherical particles \footnote{The problem is significantly simplified for spherical particles due to the symmetry properties of the sphere.}. While these models can give reasonable estimates of particle sizes from measured CLD data, if appropriate approaches are used for solving corresponding inverse problems, they are not suitable for particles whose shape deviate significantly from spherical. Even though there has been some progress in retrieving size and shape information from CLD data for populations of particles with different degrees of variation from spherical \cite{Ruf2000,Worlitschek2005,Li2005n1,Li2005n2,Noelle2006,Nere2007}, there has been no previous attempt (although Czapla et al. \cite{Czapla2010} calculated the CLD of needle shaped particles using a numerical model, the inverse problem was not solved) to obtain size and shape information for populations of needle shaped particles which are commonly present in pharmaceutical manufacturing. This is despite the fact that there are suitable geometrical models \cite{Li2005n1,Vaccaro2006} available in the literature which can be used to obtain useful  size and shape information for needle shaped particles from experimental CLD data. 

In this paper, we present an algorithm for estimating of size and shape information for needle shaped particles from experimental CLD data. We use 2 D geometrical CLD models available in the literature which are suitable for opaque particles. However, the method presented here can be extended to different 2 D and 3 D geometrical and optical CLD models for particles of arbitrary shape and optical properties. Such models would need to account for possible discontinuities along the particles' boundaries if the particles' boundaries contain strong concavities (for example the case of particle clusters). More general models would also need to account for the optical properties of the particles if the particles are not opaque.

 The optimum size range of particles in a population providing the best fit with the experimental CLD data can be directly determined by the algorithm in the case when no further  information is available, although any external information on particle size range or shape can be utilised in the algorithm as needed. We compare results from our calculations with data obtained by dynamic image analysis and laser diffraction in order to assess suitability and validity of models used.
 \begin{figure}[tbh]
\centerline{\includegraphics[width=\textwidth]{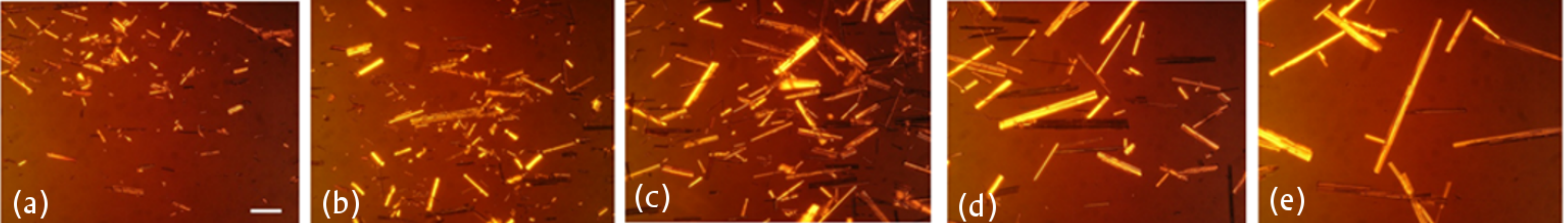}}
\caption{Microscope images (magnification factor of $\times 150$) of samples of COA after undergoing different drying conditions in the vacuum agitated drier \cite{Hamilton2012}. The samples in (a) to (e) are labelled Sample 1 to Sample 5 in Figs. \ref{fig2} and \ref{fig3}. The white horizontal line on the bottom right of (a) indicates a length of $100\mu$m. Reproduced by permission of The Royal Society of Chemistry (\href{http://doi.org/10.1039/C1AN15836H}{View Online}).}
\label{fig1}
 \end{figure}
 \begin{figure}[tbh]
\centerline{\includegraphics[width=\textwidth]{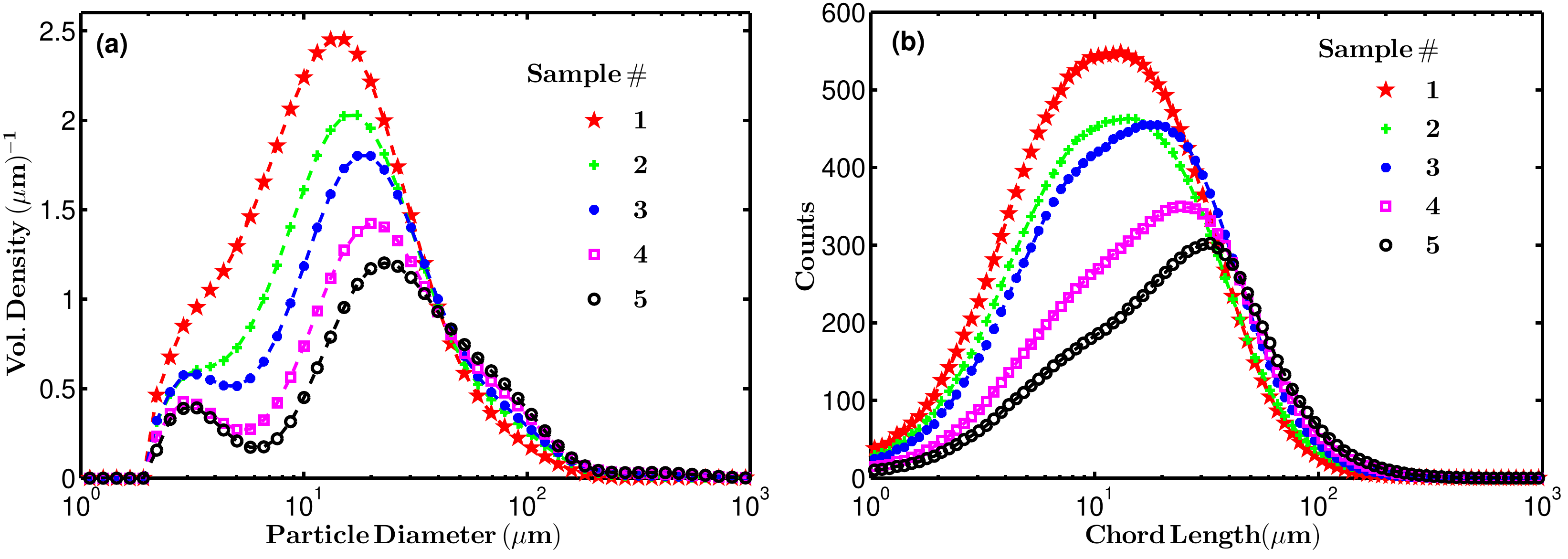}}
\caption{(a)\,Volume based particle size distribution obtained with the Malver Mastersizer and (b)\,unweighted number based chord length distribution from the FBRM probe for the samples in Fig. \ref{fig1}.}
\label{fig2}
 \end{figure}
 \begin{figure}[tbh]
\centerline{\includegraphics[width=\textwidth]{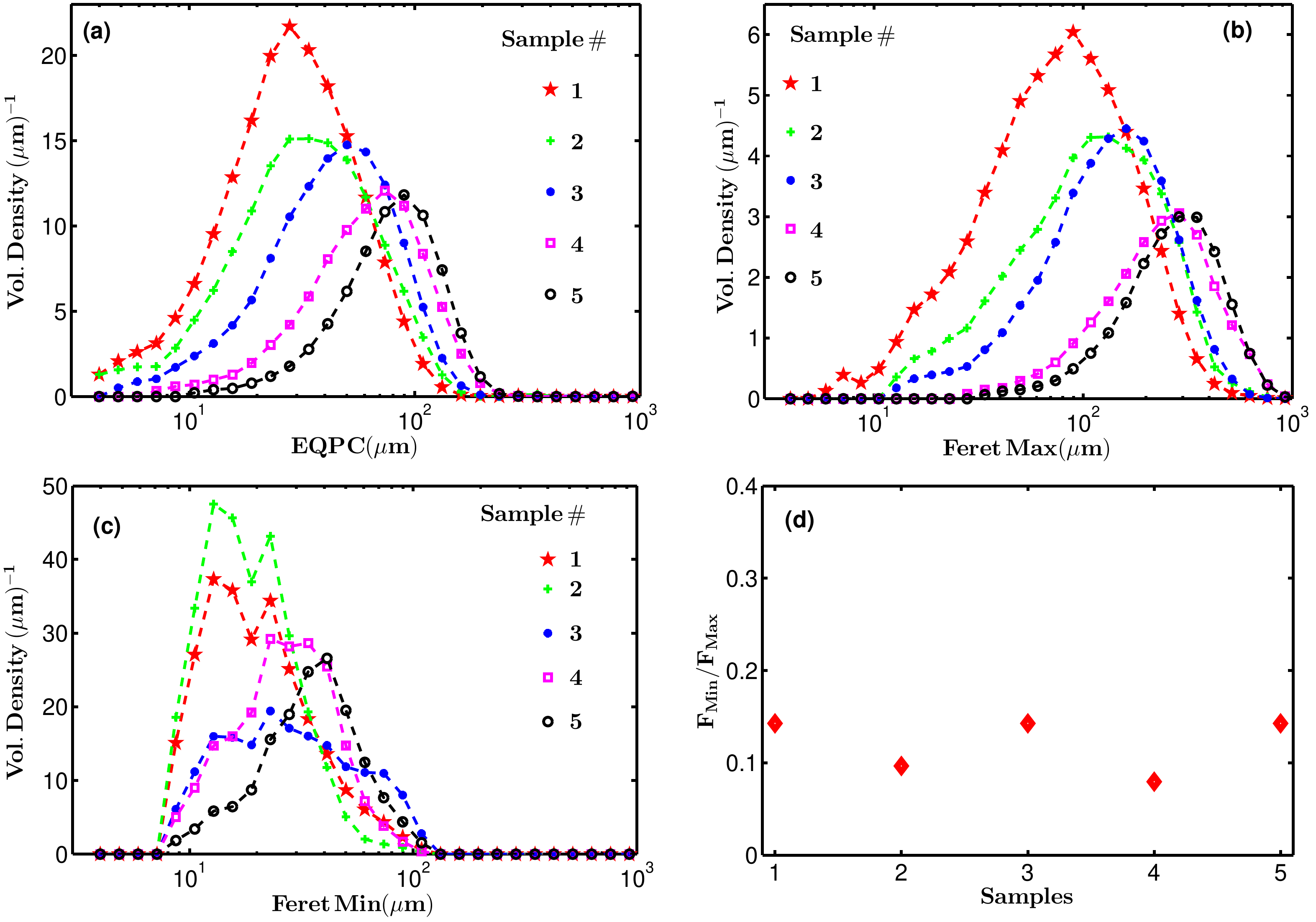}}
\caption{(a)\,Volume based EQPC diameter, maximum Feret diameter (b), and minimum Feret diameter (c) obtained by dynamic image analysis for the samples in Fig. \ref{fig1}. (d)\,A measure of the degree of elongation (aspect ratio) of the needles in Fig. \ref{fig1}.}
\label{fig3}
 \end{figure}

\section{Experimental Data}
\label{mat}

For the purpose of demonstrating and validating our technique, we shall apply the method (to be described in subsequent sections) to data obtained in a previous study \cite{Hamilton2012}. Five samples (sample 1 to sample 5) of needle-shaped particles of cellobiose octaacetate (COA) that had been subjected to different drying conditions \cite{Hamilton2011} were analysed by laser diffraction, FBRM and dynamic image analysis. The drying conditions used caused different degrees of particle attrition as shown in Fig. \ref{fig1}. Samples were dispersed in $0.1\%$ Tween 80 (Sigma-Aldrich, UK) solution in water for all particle size measurements. Laser diffraction measurements were carried out using a Malvern Mastersizer 2000 (Malvern Instruments, UK). FBRM data were obtained using a Lasentec FBRM PI-12/206 probe. Dynamic image analysis was carried out using a QICPIC (Sympatec Ltd., UK) instrument with a LIXELL wet dispersion unit. Further experimental details for the particle size analysis techniques employed can be found in the previous study \cite{Hamilton2012}.

The particle size distribution (volume weighted) estimated by laser diffraction, which assumes that the particles are spherical, for samples 1 to 5 is shown in Fig. \ref{fig2}(a). The CLD data obtained by FBRM  for the five samples is shown in Fig. \ref{fig2}(b). The equivalent projected circle EQPC diameter (which is the diameter of a circle  of equal area to the 2 D projection of a particle) distribution obtained by dynamic image analysis is shown in Fig. \ref{fig3}(a). The maximum Feret diameter (Feret Max)\footnote{See section 7 of the supplementary information and  \cite{Hamilton2012}for further description of the concepts of Feret diameter and EQPC.} obtained using dynamic image analysis, which  was shown to be a good indicator of needle length \cite{Hamilton2012}, is shown in Fig. \ref{fig3}(b). In addition, the Feret Min diameter (Feret Min) which is an indication of needle width is shown in Fig. \ref{fig3}(c). The degree of elongation (aspect ratio) of the needles can be estimated by computing the ratio of the modes of the Feret Min distributions to the modes of the Feret Max distributions. The result ($F_{min}/F_{max}$) of this calculation is shown in Fig. \ref{fig3}(d).
 The data in Figs. \ref{fig2} and \ref{fig3} will be used to compare against estimated PSDs and aspect ratios obtained from CLDs data in Fig. \ref{fig2}(b) using the algorithm described in section \ref{invalg}.

\section{Modelling Chord Length Distribution}
\label{fbrm}

The FBRM technology involves a laser beam which is focused onto a spot by a system of lenses. The focus spot is located near a sapphire window and it is rotated along a circular path at a speed of about 2ms$^{-1}$ \cite{Heinrich2012,Kail2007,Kail2008,Worlitschek2003}. The assembly of lenses is enclosed in a tubular probe which is inserted into a slurry of dispersed particles. Particles passing near the probe window reflect light back into the probe which is then detected. It is assumed that the particles are much smaller than the diameter of the circular trajectory of the laser beam, and the particles move much more slowly than the speed of the laser spot \cite{Heinrich2012}. Hence the length of arc (taken to be a straight line) made by the laser spot on a particle from which light is back scattered is just a product of the speed of the laser spot and the duration of reflection \cite{Heinrich2012}, and the corresponding chord length is recorded. Since the beam does not always pass through the centre of the particle, a range of chord lengths is recorded as a given particle encounters the beam multiple times. The FBRM device accumulates chord lengths across different particles present in the slurry for a duration pre-set by the user, after which it reports a chord length histogram, and  this data is referred to as chord length distribution (CLD).

\subsection{Calculating CLD from PSD}
\label{mod}

The CLD and PSD are related to each other and the CLD obtained from a given particle depends on both its size and shape. This size and shape information is expressed in a kernel function $A(D,L)$ which defines the CLD of a single particle of characteristic size $D$. In a population of particles, the probability of a particle being detected is linearly proportional to  its characteristic size \cite{Hobbel1991,Simmons1999,Ruf2000,Worlitschek2005,Vaccaro2006} (see also section 3 of the supplementary information). Hence the kernel needs to be weighted by the characteristic sizes of the particles in the population. The characteristic size of each particle is a monotonic function of some length scale associated with the particle \cite{Worlitschek2005}, this function depends on the shape of the particle \cite{Vaccaro2006}. For example, in the case of a population of spherical particles of different sizes the characteristic size is $D=2a_s$, where $a_s$ is the radius of a sphere. Thus the relationship between the CLD and PSD can be written as \cite{Hobbel1991}
\begin{equation}
C(L) = \int_0^{\infty}A(D,L)DX(D)dD,
\label{eq1}
\end{equation}
where $C$ is the CLD of the particle population, $L$ is chord length and $X$ is the PSD expressed as a normalised number distribution.
Equation \eqref{eq1} can be discretised and written in matrix form as \cite{Li2005n1,Hobbel1991}
\begin{equation}
\mathbf{C} = \mathbf{A\tilde{X}},
\label{eq2}
\end{equation}
 where $\mathbf{A}$ is a transformation matrix. The column vector $\mathbf{C}$ is the chord length histogram or CLD, while the column vector $\mathbf{\tilde{X}}$ is defined as 
 \begin{equation}
 \tilde{X}_i = \overline{D}_iX_i, \quad i = 1,2,3,\ldots ,N,
 \label{eq3}
 \end{equation}
 where $\mathbf{\overline{D}}$ is the vector of characteristic sizes and $\mathbf{X}$ is the unknown PSD. The characteristic sizes $D_i$ make up the bin boundaries of the PSD $X_i$, and the characteristic size of the particles bounded by the bin boundaries $D_i$ and $D_{i+1}$ is given as $\overline{D}_i=\sqrt{D_iD_{i+1}}$.
  Equation \eqref{eq2} can be rearranged so that each component of $\mathbf{\overline{D}}$ multiplies a column of $\mathbf{A}$ to give
  \begin{equation}
  \mathbf{C} = \mathbf{\tilde{A}X},
  \label{eq4}
  \end{equation}
 where
  \begin{equation}
  \tilde{A}_j = [a_{j,1}\overline{D}_1~ a_{j,2}\overline{D}_2~ \ldots~ a_{j,i}\overline{D}_i~ \ldots~ a_{j,N}\overline{D}_N], 
  \label{eq5}
  \end{equation}
 represents column $j$ of $\mathbf{\tilde{A}}$. 
   
  The matrix $\mathbf{A}$ is of dimension $M\times N$, where $M$ is the number of chord length bins in the histogram $\mathbf{C}$ and $N$ is the number of particle size bins in the histogram $\mathbf{X}$ \cite{Li2005n1}. The columns of matrix $\mathbf{A}$ are constructed as \cite{Li2005n1}
 \begin{equation}
 A_j = [a_{j,1}~ a_{j,2}~ \ldots~ a_{j,i}~ \ldots~ a_{j,N}], 
 \label{eq6}
 \end{equation}
 where
  \begin{equation}
  a_{j,i} = p_{\overline{D}_i}(L_j,L_{j+1}) 
  \label{eq7}
  \end{equation}
  is the probability that the length of a measured chord from a particle of characteristic size $\overline{D}_i$ lies between $L_j$ and $L_{j+1}$. The probabilities $p_{\overline{D}_i}(L_j,L_{j+1})$ for different particle sizes and chord length bins are calculated from appropriate probability density functions (PDF). The PDFs employed in this work are those given by the Vaccaro-Sefcik-Morbidelli (VSM) \cite{Vaccaro2006} model and the Li-Wilkinson (LW) model \cite{Li2005n1}. 
  
  The forward problem of calculating the CLD from a known PSD using Eq. \eqref{eq4} is trivial as it is mere matrix multiplication. However, the inverse problem of calculating the PSD from a known CLD is non trivial. The solution vector $\mathbf{X}$ must meet the requirement of non negativity, hence different techniques have been used in the past \cite{Worlitschek2005,Li2005n1} to fulfil this requirement. There could also be errors in the solution vector $\mathbf{X}$ if the transformation matrix $\mathbf{A}$ is inaccurate. The accuracy of the matrix $\mathbf{A}$ depends on the particle size range and the model used in calculating the probabilities in Eq. \eqref{eq7}. Here we shall describe a technique to select the most appropriate particle size range. The method employed here also guarantees the non negativity requirement of the solution vector $\mathbf{X}$. Appropriate models then need to be chosen based on any available information about the overall particle shape. In the case of needle-like particles considered here, we can use two analytical models available in the literature as discussed below. 
  
\subsection{The VSM model}

The microscope images in Fig. \ref{fig1} suggest that the shape of the particles could be represented by thin cylinders. The 2 D projections of these thin cylinders will look like the shapes in Fig. \ref{fig1}. The cylindrical VSM model \cite{Vaccaro2006} gives a PDF $X_p^c$ which defines the relative likelihood that a chord taken from a cylindrical particle has a length between $L$ and $L+dL$. To this end, the model considers all possible 3 D orientations of each cylindrical particle and calculates chord lengths from each 2 D projection. The characteristic size of a cylinder is calculated by equating to the diameter of a sphere of equivalent volume. For a thin cylinder of height $a_c$, base radius $b_c$, aspect ratio $r_c = b_c/a_c$ and characteristic size $D_c = a_c\sqrt[3]{3r_c^2/2}$, the VSM model gives the probability $X_p^c$ (for $b_c/a_c \ll 1$) as \cite{Vaccaro2006}
        \begin{equation}
       a^{\ast}X_p^c (L) = \begin{cases}
       \frac{1}{2}\frac{L}{\sqrt{r_c^2a_c^2-L^2}}\left(1 - \sqrt{1 - r_c^2}\right), & \forall L \in [0,r_ca_c[ \\
       \frac{1}{\pi}\frac{r_c^2}{\sqrt{1 - \left(\frac{L}{a_c}\right)^2}} + \frac{1}{2\pi}\frac{a_c}{L}\frac{\frac{L}{a_c}\sqrt{1 - \left(\frac{L}{a_c}\right)^2}+\cos^{-1}\left(\frac{L}{a_c}\right)}{\frac{L}{r_ca_c}\sqrt{\left(\frac{L}{r_ca_c}\right)^2 - 1}} & \forall L \in ]r_ca_c, a_c[ \\
       0 & \forall L \in [a_c, \infty[, \end{cases} 
        \label{eq8}
        \end{equation}
 where 
  \begin{equation}
        a^{\ast} = \frac{a_c}{4} + \frac{1}{2}r_ca_c\left[1 - \sqrt{1 - r_c^2} + \frac{1}{2}r_c\left(1 - \frac{4}{\pi}\sin^{-1}(r_c)\right)\right]
         \label{eq9}
         \end{equation}
         is a normalisation factor. Then the probability that the length of a measured chord from a particle of size $D_c$ falls in the bin bounded by $L_j$ and $L_{j+1}$ is calculated as 
           \begin{equation}
                  p_{c\overline{D}_i}(L_j,L_{j+1}) = \int_{L_j}^{L_{j+1}} X_p^c(L)dL. 
                  \label{eq10}
                  \end{equation}
   The integration in Eq. \eqref{eq10} is performed numerically.
   
   \subsection{The LW model}
     
   In this case, we approximate the shape of the needles in Fig. \ref{fig1} by thin ellipsoids. The model considers 2 D projections of each of ellipsoid with its major and minor axes parallel to the projection plane, so that all projections will be an ellipse of semi major axis length $a_e$, semi minor axis length $b_e$ and aspect ratio $r_e=b_e/a_e$. The length of a chord on this ellipse depends on the angle $\alpha$ between the chord and the $x$ axis (where the projection plane is the $x-y$ plane) \cite{Li2005n1}. Hence the PDF for such an ellipse is angular dependent. The PDF for different values of $\alpha$ are given by the LW model as \cite{Li2005n1}: \\
   for $\alpha = 0$ or $\pi$
      \begin{equation}
      p_{e\overline{D}_i}(L_{j,\alpha},L_{j+1,\alpha}) = \begin{cases}
      \sqrt{1 - \left(\frac{L_j}{2a_{ei}} \right)^2} - \sqrt{1 - \left(\frac{L_{j+1}}{2a_{ei}} \right)^2}, & \textrm{for}~ L_j < L_{j+1} \leq 2a_{ei} \\
      \sqrt{1 - \left(\frac{L_j}{2a_{ei}}\right)^2}, & \textrm{for}~ L_j \leq 2a_{ei} < L_{j+1} \\
      0, & \textrm{for}~ 2a_{ei} < L_j < L_{j+1}, \end{cases} 
      \label{eq11}
      \end{equation}
    for $\alpha = \pi/2$ or $3\pi/2$
       \begin{equation}
       p_{e\overline{D}_i}(L_{j,\alpha},L_{j+1,\alpha}) = \begin{cases}
       \sqrt{1 - \left(\frac{L_j}{2r_ea_{ei}} \right)^2} - \sqrt{1 - \left(\frac{L_{j+1}}{2r_ea_{ei}} \right)^2}, & \textrm{for}~ L_j < L_{j+1} \leq 2r_ea_{ei} \\
       \sqrt{1 - \left(\frac{L_j}{2r_ea_{ei}}\right)^2}, & \textrm{for}~ L_j \leq 2r_ea_{ei} < L_{j+1} \\
       0, & \textrm{for}~ 2r_ea_{ei} < L_j < L_{j+1}, \end{cases} 
       \label{eq12}
       \end{equation}
       for other values of $\alpha$
           \begin{equation}
           p_{e\overline{D}_i}(L_{j,\alpha},L_{j+1,\alpha}) = \begin{cases}
           \sqrt{1 - \frac{r_e^2 + s^2}{1 + s^2}\left(\frac{L_j}{2r_ea_{ei}} \right)^2} & \\
           - \sqrt{1 - \frac{r_e^2 + s^2}{1 + s^2}\left(\frac{L_{j+1}}{2r_ea_{ei}} \right)^2}, & \textrm{for}~ L_j < L_{j+1} \leq 2r_ea_{ei}\sqrt{\frac{1 + s^2}{r_e^2 + s^2}} \\
           \sqrt{1 - \frac{r_e^2 + s^2}{1 + s^2}\left(\frac{L_j}{2r_ea_{ei}}\right)^2}, & \textrm{for}~ L_j \leq 2r_ea_{ei}\sqrt{\frac{1+s^2}{r_e^2+s^2}} < L_{j+1} \\
           0, & \textrm{for}~ 2r_ea_{ei}\sqrt{\frac{1+s^2}{r_e^2+s^2}} < L_j < L_{j+1}, \end{cases} 
           \label{eq13}
           \end{equation}
    where $s=\tan{(\alpha)}$. The angle independent PDF is then given as
     \begin{equation}
            p_{e\overline{D}_i}(L_j,L_{j+1}) = \frac{1}{2\pi}\int_0^{2\pi} p_{e\overline{D}_i}(L_{j,\alpha},L_{j+1,\alpha})d\alpha.
            \label{eq14}
            \end{equation}
    Equation \eqref{eq14} allows the construction of the transformation matrix $\mathbf{A}$ in Eq. \eqref{eq2} which can be converted to the matrix $\tilde{\mathbf{A}}$ as described in Eq. \eqref{eq5}. The matrix $\tilde{\mathbf{A}}$ is then used to solve the inverse problem.
    
    The LW model constructs the PDF of an ellipsoidal particle by considering only one 2 D projection of the ellipsoid where the major axis is parallel to the projection plane. Hence the monotonic function which gives the characteristic size $D_e$ of the resulting ellipse is obtained from the area of a circle of equivalent area. Hence, using $r_e = b_e/a_e$, the characteristic size is given as $D_e=2a_e\sqrt{r_e}$.
    
     \begin{figure}[tbh]
    \centerline{\includegraphics[width=\textwidth]{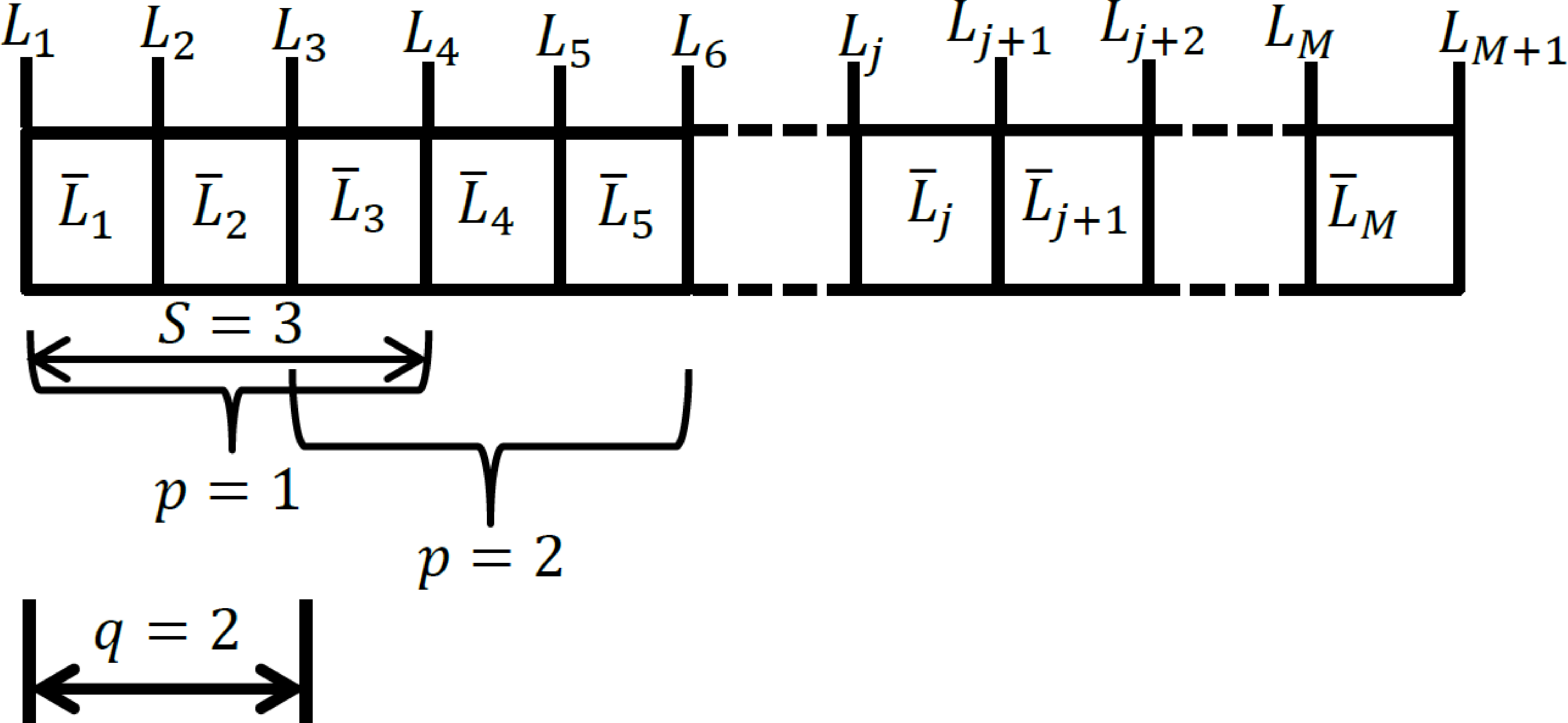}}
    \caption{Pictorial representation of the bins and bin boundaries of the CLD histogram showing a window of size $S$ at the first two positions set by $p=1$ and $p=2$ shifted by $q$. The window is moved in such a way that some of the bins contained in the window at $p=1$ overlap some of the bins of the window at $p=2$.}
    \label{fig4}
     \end{figure}
      
\section{Inversion Algorithm}
\label{invalg}

As mentioned in the introduction, one important factor which can be used to constrain inverse problem solutions is the size range ($D_{min}$ to $D_{max}$) of particles used in the calculations, where $D_{min}$ is the smallest particle size and $D_{max}$ is the largest particle size in the population. Since this information is not always readily available, we introduce an inversion algorithm which is capable of automatically determining the best values of $D_{min}$ and $D_{max}$ to solve the inverse problem. We use the bin boundaries of the chord length histogram to specify the size range boundaries $D_{min}$ and $D_{max}$. A number $S$ of consecutive bins of the chord length histogram are chosen, these bins make up a window of width $S$. This means that the width (or size) of a window is the number of bins contained within that window.  The geometric mean of the first two bin boundaries of a window is taken as $D_{min}$ and the geometric mean of the last two bins of a window is taken as $D_{max}$. The procedure is outlined below.

The boundaries of the chord length histogram are labelled as
 \begin{equation}
 L_j, ~ j=1,2,3, \ldots, M+1
 \label{eq15}
 \end{equation}
 as illustrated in Fig. \ref{fig4}. The characteristic chord length $\overline{L}_j$ of bin $j$ is the geometric mean of the chord lengths of its boundaries
  \begin{equation}
  \overline{L}_j = \sqrt{L_iL_{j+1}}.
  \label{eq16}
  \end{equation}
  At the beginning of the calculation, the first $w$ bins of the chord length histogram are chosen, so that $S = w$, $D_{min} = \overline{L}_1$ and $D_{max} = \overline{L}_w$. After the first iteration (see steps 5 to 9 in the algorithm below), a new set of bin boundaries are selected. This new set of bin boundaries is made up of the same number of bins $S$ as the previous set, but it is shifted to the right of the previous set by an amount $q$. That is, there are $q$ bins between the beginning of the first set of bins and the beginning of the second set. The shift is made in such a way that the two set of bins overlap each other (that is, $q<S$). For example, in the case illustrated in Fig. \ref{fig4}, the window initially runs from bin boundary $L_1$ to bin boundary $L_4$. At this position, the window contains bins $\overline{L}_1$ to $\overline{L}_3$ so that the width of the window is $S = 3$.  At the end of the first iteration, a new set of bins are chosen, this time starting from bin boundary $L_3$ and ending at bin boundary $L_6$ as in Fig. \ref{fig4}. The number of bins in the new set of bins (or window) is the same as before $S = 3$. Each window (or set of bins) is identified by its position index $p$. In the case shown in Fig. \ref{fig4}, the value of the first position index is $p = 1$ and the value of the second position index is $p = 2$. There are two bins between the beginning of the window at $p = 1$ and the beginning of the window at $p = 2$ so that $q = 2 < S$. At the end of the second iteration, the window is shifted to the right again, while maintaining fixed values of $S$ and $q$. This process continues until the last bin boundary of the chord length histogram is reached. 
  
  Each time a set of bins are chosen, the values of $D_{min}$ and $D_{max}$ are calculated as
   \begin{subequations}
   \begin{align}
   D_{min} & =  \overline{L}_1\beta^{(p-1)q} \\
   D_{max} & = D_{min}\beta^{\left(S-1\right)},
   \end{align}
   \label{eq17}
   \end{subequations}
where $\beta = \overline{L}_{j+1}/\overline{L}_j$. The position index of the windows take values  
   \begin{equation}
  p = 1,2,3, \ldots, \left\lfloor \frac{M}{q} \right\rfloor
   \label{eq18},
   \end{equation}
  where the floor function $\lfloor \cdot \rfloor$  returns the value of the largest integer that is less than or equal to $M/q$.

Once the values of $D_{min}$ and $D_{max}$ have been calculated from Eq. \eqref{eq17}, then particle size bins are constructed. The bin boundaries $D_i$ of the particle size bins are calculated as 
   \begin{equation}
 D_i = D_{min}\mu^{i-1}, \quad i=1,2, \ldots , N+1
   \label{eq19}
   \end{equation}
   where
      \begin{equation}
     \mu = \left(\frac{D_{max}}{D_{min}}\right)^{\frac{1}{N}}
      \label{eq20},
      \end{equation}
   where $N$ is the chosen number of particle size bins. The characteristic size of a particle size bin is calculated as
      \begin{equation}
     \overline{D}_i = \sqrt{D_iD_{i+1}}.
      \label{eq21}
      \end{equation}
Once the characteristic particle sizes $[\overline{D}_1,\overline{D}_N]$ have been constructed, then the transformation matrix $\tilde{\mathbf{A}}$ can be constructed (for a chosen aspect ratio) as in Eq. \eqref{eq5}.
The chord lengths reported by the FBRM sensor run from $1\mu$m to $1000\mu$m. However, the particle size range $[\overline{D}_1,\overline{D}_N]$ set by a window will not necessarily cover the entire size range of $1\mu$m to $1000\mu$m. To account for the other sizes that may not be covered by a window, the length weighted transformation matrix $\tilde{\mathbf{A}}$ is augmented with columns of ones as appropriate. Then the particle sizes are extended to the left of $\overline{D}_1$ down to $1\mu$m and to the right of $\overline{D}_N$ up to $1000\mu$m as appropriate. This ensures that the recovered PSD covers the entire particle sizes from $1\mu$m to $1000\mu$m. The process of augmenting the transformation matrix with columns of ones corresponds to the addition of slack variables in an optimisation problem \cite{Boyd2004} (see also  section 1 of supplementary information).

To guarantee non negative PSD the vector $\mathbf{X}$ is written as \cite{Kaasaleinen2001}
 \begin{equation}
X_i = e^{\gamma_i},~i = 1,2,3, \ldots, N,
 \label{eq22}
 \end{equation}
 where $\gamma_i$ are arbitrary fitting parameters.
Then Eq. \eqref{eq4} is rewritten as 
 \begin{equation}
\mathbf{C} = \mathbf{\tilde{A}X}+\mbox{\boldmath{$\epsilon$}},
 \label{eq23}
 \end{equation}
where $\mbox{\boldmath{$\epsilon$}}$ is an additive error between the model prediction and the actual measurement. The vector $\mathbf{X}(r)$ at the chosen aspect ratio $r$ is then obtained by searching for $\gamma_i$ which minimises the objective function $f_1$ given as\footnote{In all the calculations here a value of $N = 70$ was used for both VSM amd LW models (section 2 of the supplementary information).}
 \begin{equation}
f_1 = \sum_{j=1}^{M}{\left[C_j^{\ast} - \sum_{i=1}^N{\tilde{A}_{ji}X_i}\right]^2},
 \label{eq24}
 \end{equation}
 where $C_j^{\ast}$ is the experimentally measured CLD.
This nonlinear least squares problem was solved with the Levenberg-Marquardt (LM) algorithm (implemented in Matlab in this work). Then starting with an initial value\footnote{Different choices of initial $\gamma_i$ resulted in the same optimum solution.} for the vector $\gamma_i$ the LM algorithm performs a successive iteration until an optimum $\gamma_i$ is reached. The iterations are terminated when a specified tolerance in the difference between successive function evaluations is reached. In this case we used a tolerance of $10^{-6}$ since the results did not change for values of tolerance below $10^{-4}$. An initial value of $\mbox{\boldmath{$\gamma = 0$}}$ was used in the LM algorithm.

\begin{figure}[tbh]
    \centerline{\includegraphics[width=0.5\textwidth]{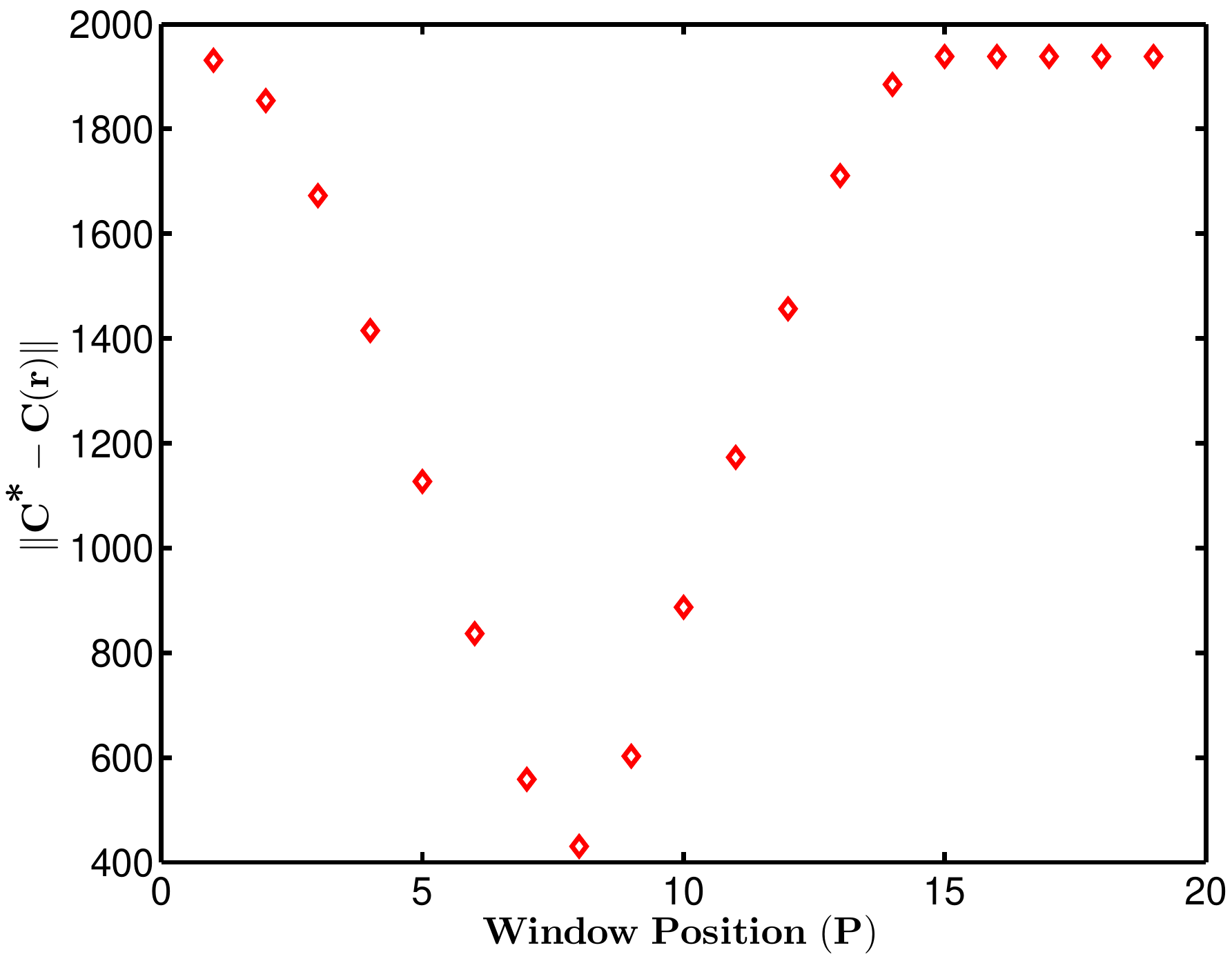}}
    \caption{An example of the minimisation of the $L_2$ norm in Eq. \eqref{eq25} when a window of a given size approaches and passes its optimum position along the bin boundaries of the chord length histogram.}
    \label{fig5}
     \end{figure}

The solution vector $\mathbf{X}$ obtained this way (using Eq. \eqref{eq22}) is dependent on the chosen aspect ratio $r$ (hence $\mathbf{X} = \mathbf{X}(r)$), window size $S$ and window position $p$. Thus, starting with a window of a chosen size\footnote{The values $q=2$, and initial window size $S=6$ were used for both the VSM and LW models (section 2 of supplementary information).} and at position set at $p=1$, a solution vector $\mathbf{X}(r)$ is obtained for the chosen aspect ratio. Then the forward problem is solved to obtain a CLD $\mathbf{C}(r)$ at that aspect ratio and window position $p=1$. The window position is advanced one step forward and the calculation repeated until the last bin of the chord length histogram is reached. The window position at which the $L_2$ norm
 \begin{equation}
\left\|\mathbf{C}^{\ast}-\mathbf{C}(r)\right\|
 \label{eq25}
 \end{equation}
is minimized is the optimum window position for that window size. This optimum window then sets the particle size range to construct the optimum transformation matrix $\mathbf{\tilde{A}}$ at that window size. The case of $S=20$ applied to the CLD from Sample 1 (using the LW model) is shown in Fig. \ref{fig5}. The procedure is repeated using windows of different sizes and eventually the optimum window size and position which set the particle size range for the chosen aspect ratio is obtained. The whole process is repeated at different aspect ratios, and for each aspect ratio the particle size range is obtained from the optimum window size and position. 

The key parameters of the algorithm are the quantities $r$, $S$, $q$ and $N$. An extensive study (see section 2 of the supplementary information) has shown that a value of $N=70$ is suitable for the two models implemented here. The algorithm starts with an initial window size $S$ after which the window size is increased. In section 2 of the supplementary information it was demonstrated that initial values of $S$ from 2 up to 50 give consistent results for $N\gtrsim 60$. However an initial value of $S=6$ was used in all the calculations here for more accuracy. The smallest value of $q$ that can be used is $q=1$, however a value of $q=2$ was used here since there is no significant change in the level of accuracy obtained at $q=1$. The value of $q=1$ will only lead to greater resolution as can be seen in Fig. \ref{fig5}. Once the initial value of $S$, the values of $q$ and $N$ have been fixed, then the algorithm loops through subsequent values of $S$ at all desired values of $r$ as summarised below:
\begin{enumerate}
\item{Choose an aspect ratio $r$.}
\item{Choose a number $S$ of bins of the chord length histogram.}
\item{Start at window position $p=1$.}
\item{Obtain the values of $D_{min}$ and $D_{max}$ dictated by the window at the position set by $p$.}
\item{Construct matrix $\tilde{\mathbf{A}}$ corresponding to the values of $D_{min}$ and $D_{max}$ in step 4.}
\item{Augment matrix $\tilde{\mathbf{A}}$ with columns of ones and extend the particle size range as necessary.}
\item{Implement the LM algorithm to calculate $\mbox{\boldmath{$\gamma$}}$ starting with $\mbox{\boldmath{$\gamma = 0$}}$, and then calculate $\mathbf{X}(r)$ from Eq. \eqref{eq22}.}
\item{Calculate $\mathbf{C}(r)$ from Eq. \eqref{eq4}.}
\item{Calculate the $L_2$ norm in Eq. \eqref{eq25} for the given values of  $r$, $S$ and $p$.}
\item{Update $p$ and repeat steps 4 to 9 for the same values of $r$ and $S$ until the last bin of the chord length histogram is reached.}
\item{Choose the best window position (the window position with the minimum $L_2$ norm as in Fig. \ref{fig5}) for the given values of $r$ and $S$.}
\item{Update the window size $S$ and repeat steps 3 to 11.}
\item{For a given $r$ obtain the window position and size at which the $L_2$ norm in Eq. \eqref{eq25} attains its minimum. Record the particle size range corresponding to this window position and size.}
\item{Update $r$ and repeat steps 2 to 13.}
\end{enumerate}

The values of $S$ used in the algorithm will depend on the desired level of accuracy. Using closely spaced values of $S$ will result in greater accuracy but with the consequent increase in computational time. However widely spaced values of $S$ will lead to lower computational times but less accurate results. The window sizes are calculated as
   \begin{equation}
  S_k = S_0 + \left\lfloor \left(k - 1\right) \frac{M}{N_w}\right\rfloor
   \label{eq26},
   \end{equation}
where $\lfloor \cdot \rfloor$ is the floor function discussed in Eq. \eqref{eq18}, $S_0$ is the initial window size and $N_w < M$ is the desired number of windows. A value of $N_w = 50$ was used in the calculations here. The values of $r$ chosen depends on the desired range of aspect ratios to explore.  

Having obtained the optimum particle size ranges at different aspect ratios for a particular sample, then the optimum aspect ratio for that sample can be chosen using a suitable procedure. The simplest procedure would have been to pick the aspect ratio at which the $L_2$ norm reaches its global minimum. However, the simulations show (see section 6 of supplementary information) that when the number of particle size bins is large enough the $L_2$ norm in Eq. \eqref{eq25} does not show a clear global minimum. Instead it decreases with increasing aspect ratio and then levels off after some critical aspect ratio. Hence unique shape information cannot be obtained using the objective function in Eq. \eqref{eq24}. 

This problem of non uniqueness can be removed if the shape of the recovered PSD ($X_i$ in Eq. \eqref{eq24}) is taken into account. As the aspect ratio deviates further from some reasonable value for a particular sample, the recovered PSD shows an increasing degree of oscillations. These oscillations could also occur even when the aspect ratio is physically reasonable, but they become more pronounced as the aspect ratio deviates from realistic values. To address this issue, one can introduce a modified function which reduces these oscillations by minimising the total variation in the PSD. Here we use a new objective function $f_2$ given as
 \begin{equation}
f_2 = \sum_{j=1}^{M}{\left[C_j^{\ast} - \sum_{i=1}^N{\tilde{A}_{ji}X_i}\right]^2} + \lambda\sum_{i=1}^N{X_i^2},
 \label{eq27}
 \end{equation}
where the parameter $\lambda$ sets the level of the penalty function imposed on the norm of the PSD. The value of $\lambda$ is chosen by comparing the relative magnitude of the two sums of squares in Eq. \eqref{eq27} (see section 6 of supplementary information for more details). The optimum particle size ranges at different aspect ratios obtained using the inversion algorithm above are used to construct the transformation matrix $\tilde{\mathbf{A}}$ (in Eq. \eqref{eq27}) at the corresponding aspect ratios. The optimum aspect ratio is chosen as the value of $r$ at which the objective function $f_2$ reaches its global minimum for a carefully chosen value of $\lambda$. The corresponding PSD at which $f_2$ reaches its global minimum is then chosen as the optimum PSD.

For a meaningful comparison of calculated PSD with experimentally measured PSD from laser diffraction and imaging, it is necessary that the calculated PSD be cast as a volume based distribution. This is because some instruments report PSD in terms of a volume based distribution for example Figs. \ref{fig3}(a), \ref{fig3}(b) and \ref{fig3}(c). The volume based PSD $\mathbf{X}^v$ given by \cite{Holdich2002}
      \begin{equation}
   X^v_i = \frac{X^o_i\overline{D}^3_i}{\sum_i^N{X^o_i\overline{D}^3_i}},
      \label{eq28}
      \end{equation}
(where $\mathbf{X}^o$ is the optimum number based PSD which minimises the objective function $f_2$ in Eq. \eqref{eq27}) could lead to artificial peaks at large particle sizes if there are small fluctuations in the right hand tail of the number based PSD estimates (see section 5 of supplementary information). These fluctuations are usually very small with an amplitude of the order of $0.1\%$ of the peak of the number based PSD $X_i$ in Eq. \eqref{eq27}. Because the amplitude of the fluctuation is small, they are not removed by the penalty function in Eq. \eqref{eq27}. Also, the level of penalty imposed on the recovered number based PSD needs to be maintained at reasonable levels so that the recovered PSD does not get skewed. This situation requires that a suitable regularisation be applied to the recovered volume based PSD such as the $\lambda$ parameter in Eq. \eqref{eq27}. This problem can be addressed by restating the inverse problem as follows: \\ 
Calculate the CLD $C^o_j$ given by 
 \begin{equation}
C_j^o = \tilde{A}^o_{ji}\hat{X}^o_i,
 \label{eq29}
 \end{equation}
 where $\tilde{A}^o_{ji}$ is the optimum transformation matrix obtained by the inversion algorithm and
  \begin{equation}
\hat{X}^o_i = \frac{X^o_i}{\sum_i^N{X_i^o}}.
  \label{eq30}
  \end{equation}
  If the volume based PSD $X^v_i$ was known, then the CLD $C_j^o$ can also be calculated from
      \begin{equation}
    C^o_j = \overline{A}^o_{ji}\overline{X}^v_i,
      \label{eq31}
      \end{equation}
   where
    \begin{subequations}
    \begin{align}
    \overline{A}^o_{ji} & = \frac{\tilde{A}^o_{ji}}{\overline{D}^3_i} \\
     X^v_i & = \frac{\hat{X}^o_i\overline{D}^3_i}{\sum_i^N{\hat{X}^o_i\overline{D}^3_i}} \\
    \overline{X}^v_i & = \left[\sum_{i}^{N}{X^o_i\overline{D}^3_i}\right]X^v_i.
    \end{align}
    \label{eq32}
    \end{subequations}
  Equation \eqref{eq31} is the forward problem for the volume based PSD similar to the case of Eq. \eqref{eq4} for the number based PSD. However, since the volume based PSD is not known, then an objective function similar to $f_2$ in Eq. \eqref{eq27} can be formulated to recover the volume based PSD. This objective function $f_3$ is given as
  \begin{figure}[tbh]
   \centerline{\includegraphics[width=\textwidth]{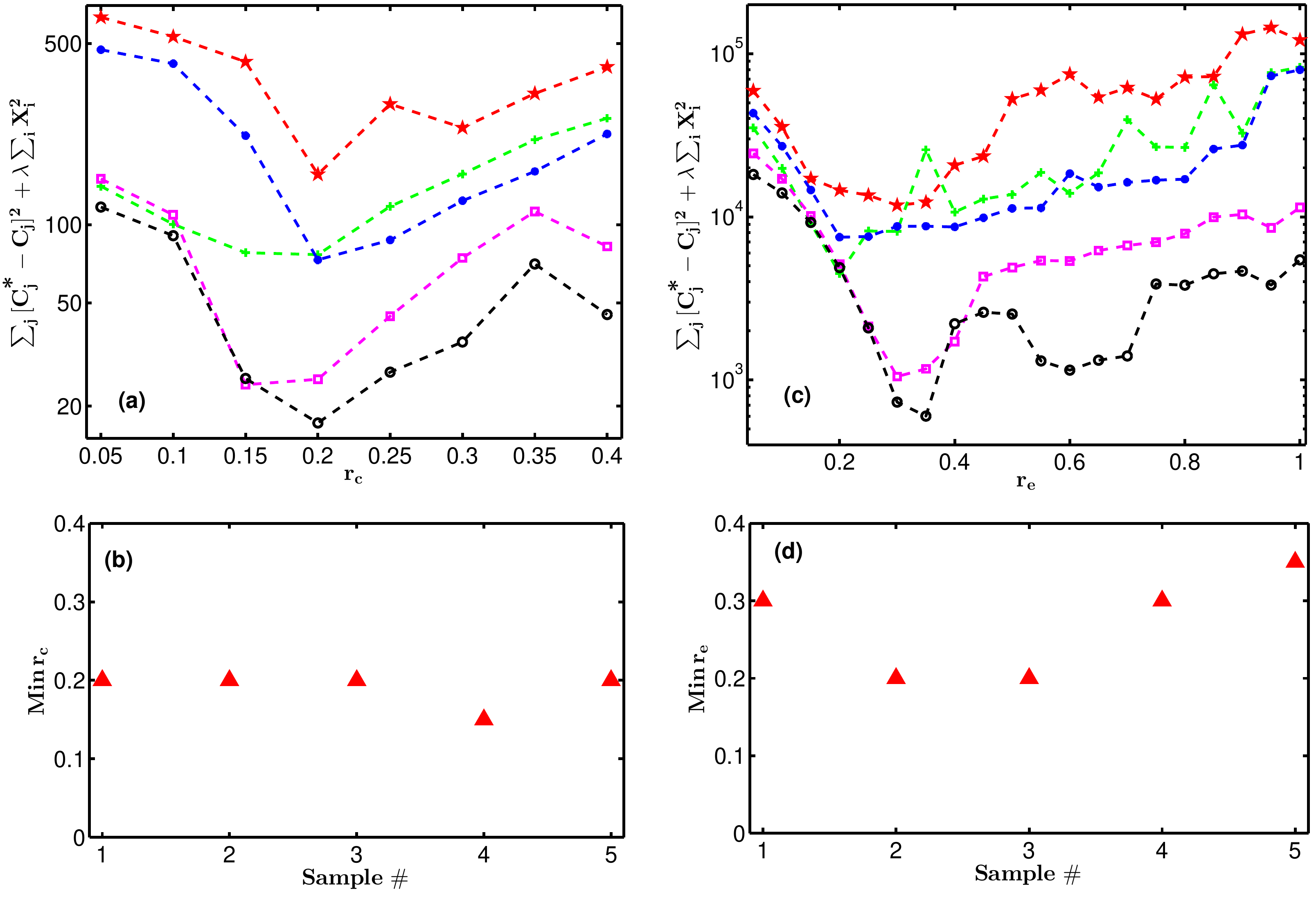}}
   \caption{(a)\,The minimum values of the objective function in Eq. \eqref{eq27} versus the aspect ratio (the minimum values of the objective function for all window sizes and positions for each sample indicated with symbols as: Sample 1 - red pentagrams, Sample 2 - green crosses, Sample 3 - blue asterisks, Sample 4 - magenta squares, Sample 5 - black circles) obtained with the VSM model. (b)\,The aspect ratios ($Min~r_c$) at which the objective function reaches a global minimum for each Sample obtained with the VSM model. (c)\,Similar to (a) obtained with the LW model. (d)\,Similar to (b) with the LW model.}
   \label{fig6}
    \end{figure}
      \begin{equation}
    f_3 = \sum_{j=1}^M{\left[C_j^o - \sum_{i=1}^N{\overline{A}^o_{ji}\overline{X}^v_i} \right]^2} + \lambda\sum_{i=1}^N{\left[\,\overline{X}_i^v\right]^2}.
      \label{eq33}
      \end{equation}
            
 This allows $\overline{X}^v_i$ (obtained to some weighting factor due to Eq. \eqref{eq32}(c)) to be calculated as
       \begin{equation}
     \overline{X}^v_i = e^{\gamma^v_i},~i = 1,2, \ldots,N,
       \label{eq34}
       \end{equation}
    where $\gamma_i^v$ is an arbitrary parameter which is used to minimise the objective function $f_3$ for a carefully chosen $\lambda$.
       The weighted volume based PSD is then normalised and made grid independent as 
              \begin{equation}
            \tilde{X}^v_i = \frac{\overline{X}^v_i}{\left(D_{i+1}-D_i\right)\sum_i^N{\overline{X}^v_i}}.
              \label{eq35}
              \end{equation}

\section{Results and Discussion}
\label{res}

Once the optimum particle size ranges at the different aspect ratios have been obtained using the inversion algorithm, then the optimum aspect ratio for each sample can be determined by selecting the aspect ratio at which the objective function $f_2$ (in Eq. \eqref{eq27}) reaches its global minimum. The objective function $f_2$ at different aspect ratios $r_c\in [0,0.4]$ for the five samples in Fig. \ref{fig1} is shown in Fig. \ref{fig6}(a) for the case of the VSM model\footnote{The values of $\lambda = 0.01$ and $\lambda = 0.2$ were used in Eq. \eqref{eq26} for the VSM and LW model respectively (section 6 of the supplementary information)}. The function $f_2$ reaches its global minimum at $r_c\approx 0.2$ as in Fig. \ref{fig6}(b). The calculations with the VSM model was restricted to the range $r_c\in [0, 0.4]$ because the thin cylindrical VSM model is only valid for $r_c \ll 1$ \cite{Vaccaro2006}. 

Figure \ref{fig6}(c) shows a similar result to Fig. \ref{fig6}(a) for the same samples in Fig. \ref{fig1} for the case of the LW model. The function $f_2$ reaches its global minimum for $r_e\approx 0.3$ as in Fig. \ref{fig6}(d). In this case, the aspect ratios $r_e$ cover a broader range $r_e\in [0, 1]$ since the LW model is valid for $r_e\in [0, 1]$. The aspect ratios predicted by the VSM and LW models in Figs. \ref{fig6}(b) and \ref{fig6}(d) are comparable to the aspect ratios estimated from image data in Fig. \ref{fig3}(d), although the calculated aspect ratios appear slightly higher. 

The aspect ratios predicted by the VSM model in Fig. \ref{fig6}(b) are closer to the estimated aspect ratios in Fig. \ref{fig3}(d) when compared with the aspect ratios predicted by the LW model in Fig. \ref{fig6}(d). This could be because the cylindrical shape used in the VSM model is closer to the shape of the particles in Fig. \ref{fig1} than the ellipsoidal shape used in the LW model. We also note that the VSM model gives much lower error norm than the LW model for the same aspect ratio as seen in Figs. \ref{fig6}(a) and \ref{fig6}(c), and this is also the case when $\lambda = 0$ (see section 6 of supplementary information). The effect of shape on the level of accuracy reached in the calculations is demonstrated by the fact that when the LW model is applied to a system of spherical particles (section 6 of supplementary information), the error norm obtained in that case is comparable to the error norm obtained when the VSM model is applied to the needle particles.
                   \begin{figure}[tbh]
                  \centerline{\includegraphics[width=\textwidth]{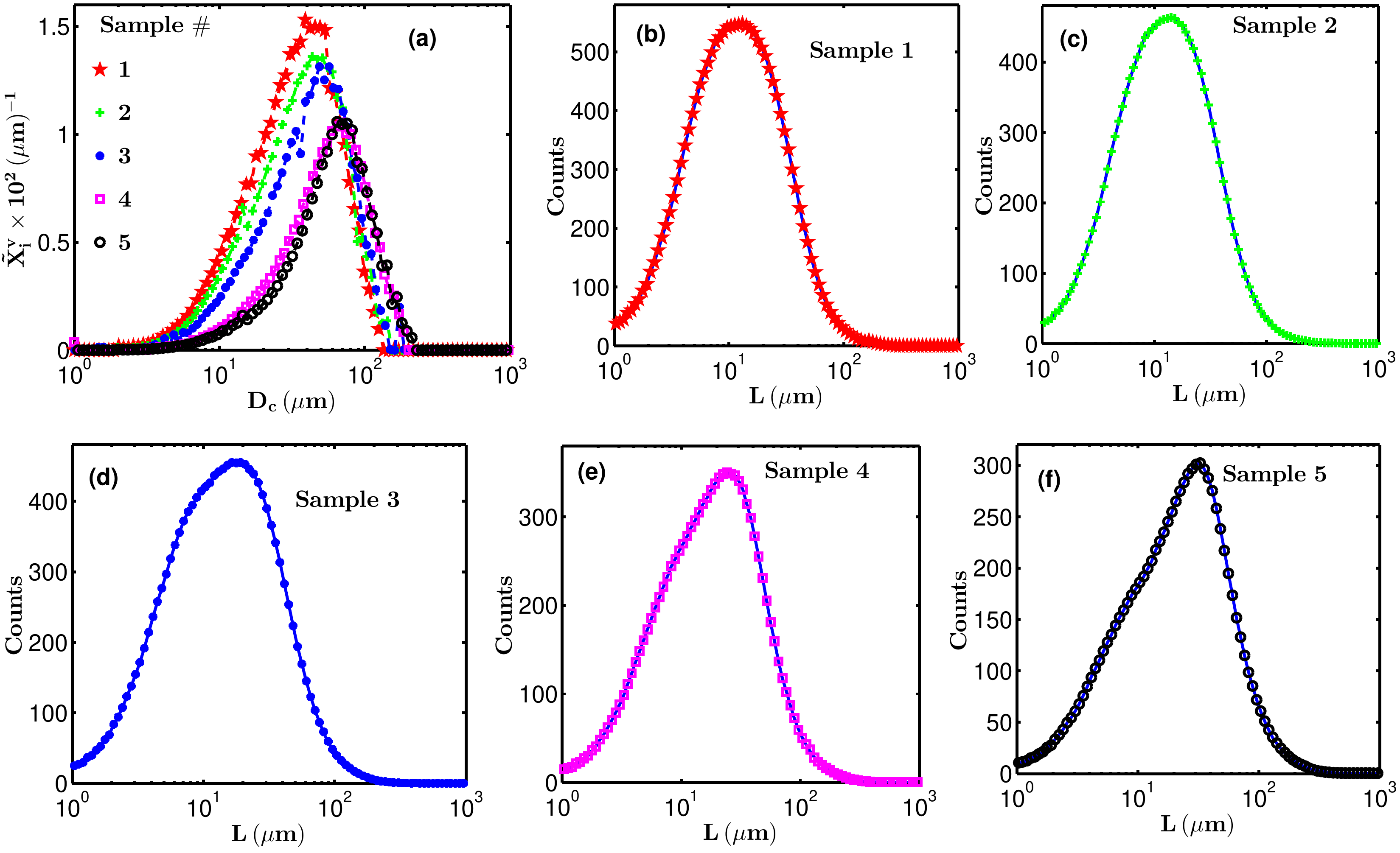}}
                  \caption{(a)\,The recovered volume based PSDs calculated from the objective function in Eq. \eqref{eq33} for $\lambda = 0$ (with the VSM model) at the minimum aspect ratios (shown in Fig. \ref{fig6}(b)) for each Sample. (b)-(f)\,Calculated (symbols) and measured (solid line) Chord Length Distributions for the Samples indicated in each Figure. The calculated CLDs were obtained by solving the forward problem in Eq. \eqref{eq4} using the number based PSD which minimise the objective function in Eq. \eqref{eq23} for $\lambda=0.01$.}
                  \label{fig7}
                   \end{figure}

Figure \ref{fig7}(a) shows the recovered volume based PSD calculated by minimising the objective function $f_3$ in Eq. \eqref{eq33} using the optimum aspect ratios in Fig. \ref{fig6}(b)\footnote{The values of $\lambda =0$ and $\lambda = 8\times 10^{-15}$ were used in Eq. \eqref{eq33} for the VSM and LW models respectively (see section 6 of the supplementary information).} for the case of the VSM model. The transformation matrix $\tilde{\mathbf{A}}^o$ used in Eq. \eqref{eq32} was constructed using the optimum particle size range obtained by the inversion algorithm and aspect ratios shown in Fig. \ref{fig6}(b). The matrix $\tilde{\mathbf{A}}^o$ is then weighted as in Eq. \eqref{eq32}(a) to obtain the matrix $\overline{\mathbf{A}}^o$. The volume based PSD $\tilde{\mathbf{X}}^v$ normalised and rescaled as in Eq. \eqref{eq35} are shown in Fig. \ref{fig7}(a). 

The PSDs in Fig. \ref{fig7}(a) are shown as a function of the characteristic particle size $D_c$. This PSD can be  compared to the data from laser diffraction in Fig. \ref{fig2}(a) and EQPC diameter in Fig. \ref{fig3}(a). The particle sizes in Fig. \ref{fig7}(a) cover a range of $D_c\approx 7\mu$m to $D_c\approx 200\mu$m. The modes of the distributions cover a range of $D_c\approx 40\mu$m to $D_c\approx 70\mu$m, with the sizes increasing from sample 1 to sample 5. This is consistent with the data from laser diffraction in Fig. \ref{fig2}(a) where the diameters cover a range of about $2\mu$m to about $200\mu$m. The modes of the distributions cover a range of about $10\mu$m to about $30\mu$m with the particle sizes increasing from sample 1 to sample 5. Similarly, the EQPC diameters in Fig. \ref{fig3}(a) cover a range of about $10\mu$m to about $200\mu$m with the modes running from about $30\mu$m to about $100\mu$m, and the sizes increasing from sample 1 to sample 5. The peaks of the PSDs from the laser diffraction in Fig. \ref{fig2}(a) and EQPC diameters in Fig. \ref{fig3}(a) decrease from sample 1 to sample 5 which is consistent with the results reported in Fig. \ref{fig7}(a).

The symbols in Figs. \ref{fig7}(b) to \ref{fig7}(f) show the calculated (using the VSM model) CLDs for the five samples in Fig. \ref{fig1}. The CLDs were calculated from  Eq. \eqref{eq4} using the number based PSD which minimises the objective function $f_2$ in Eq. \eqref{eq27}. The calculations were done at the optimum aspect ratios in Fig. \ref{fig6}(b). The blue solid lines in Figs. \ref{fig7}(b) to \ref{fig7}(f) are the experimentally measured CLDs for the five samples shown in Fig. \ref{fig2}(b). The agreement between the calculated CLDs and the experimentally measured CLDs in Figs. \ref{fig7}(b) to \ref{fig7}(f) is near perfect. This level of agreement between the calculated PSD and CLD with the experimentally measured PSD and CLD demonstrates the level of accuracy that can be achieved with this algorithm.

                   \begin{figure}[tbh]
                  \centerline{\includegraphics[width=\textwidth]{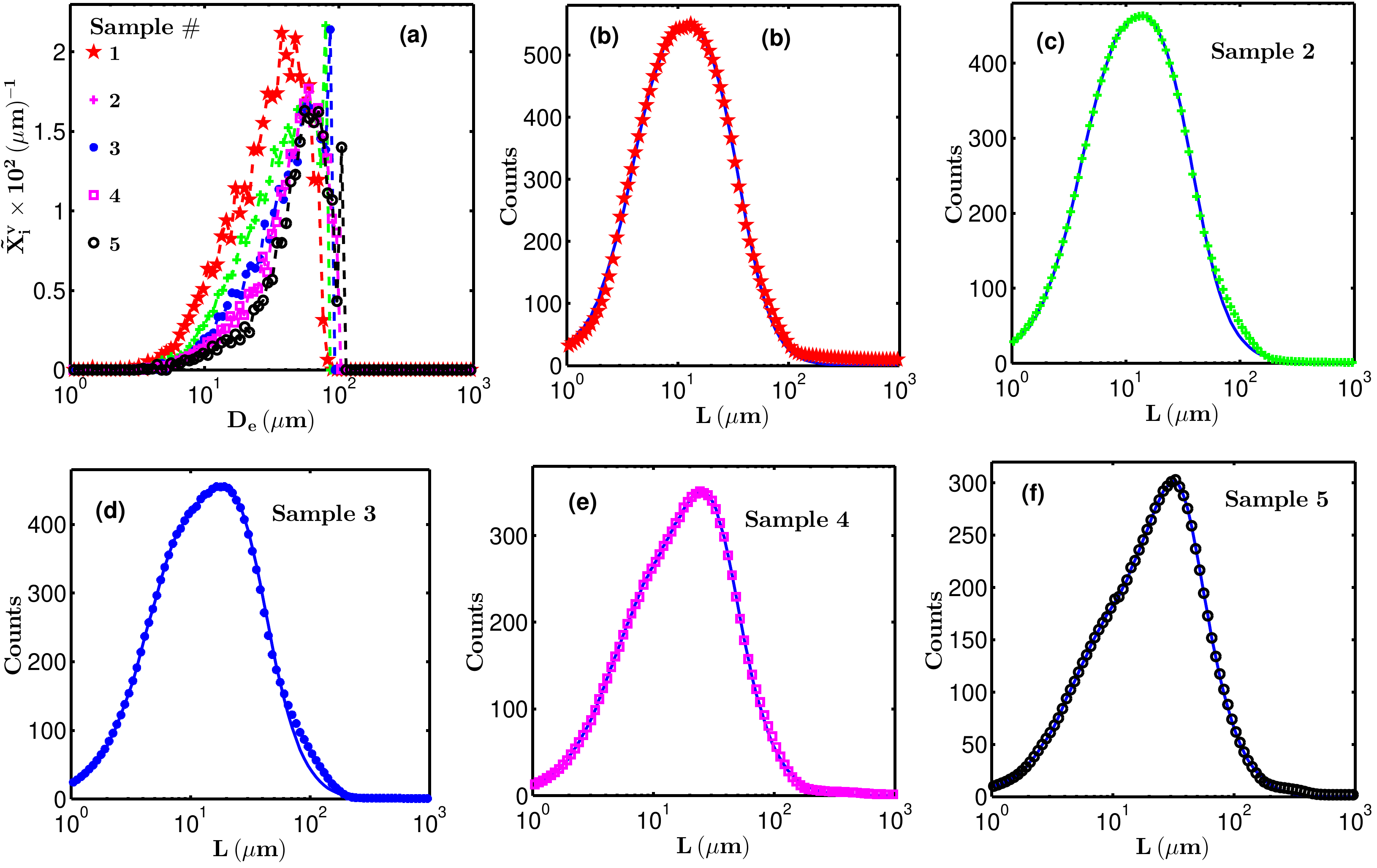}}
                  \caption{Similar to Fig. \ref{fig7} obtained with the LW model. In this case the volume weighted PSDs were obtained at $\lambda = 10^{-14}$ from Eq.\eqref{eq33}, while the CLDs correspond to number based PSD obtained at $\lambda = 0.2$ from Eq. \eqref{eq27}.}
                  \label{fig8}
                   \end{figure}
                   \begin{figure}[tbh]
                  \centerline{\includegraphics[width=\textwidth]{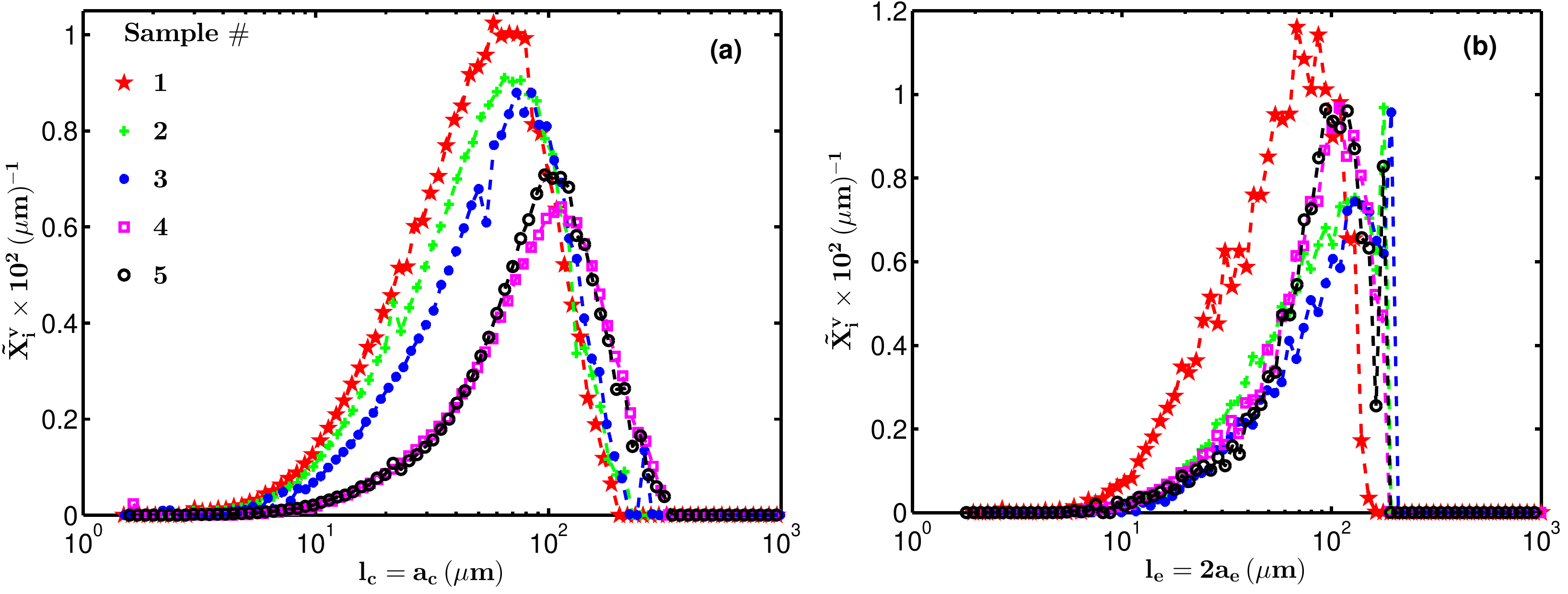}}
                  \caption{Particle lengths for the five samples calculated with (a) the VSM model and (b) the LW model.}
                  \label{fig9}
                   \end{figure}
                  
Figure \ref{fig8}(a) shows the volume based PSDs for the five samples in Fig. \ref{fig1} calculated with the LW model. The calculations were done in a similar manner as in Fig. \ref{fig7}(a). The distributions are plotted as a function of the characteristic size $D_e$ which are comparable to the laser diffraction data in Fig. \ref{fig2}(a) and EQPC data in Fig. \ref{fig3}(a). The level of consistency of the volume based PSDs in Fig. \ref{fig8}(a) to the particle sizes in Figs. \ref{fig2}(a) and \ref{fig3}(a) is similar to the case of Fig. \ref{fig7}(a). The range of particle sizes in Fig. \ref{fig8}(a) and the modes of the distributions in Fig. \ref{fig8}(a) are close to the measured data in Figs. \ref{fig2}(a) and \ref{fig3}(a). However, the calculated PSDs in Fig. \ref{fig8}(a) show some oscillations. This is also reflected in the fact that the error norms between the measured CLD and calculated CLD with the LW model is higher than the corresponding error norm of the calculations with the VSM model as seen in Figs. \ref{fig6}(a) and \ref{fig6}(c).

The symbols in Figs. \ref{fig8}(b) to \ref{fig8}(f) show the calculated (using the LW model) CLDs for the samples in Fig. \ref{fig1}. The calculations were done in a manner similar to the case of Figs. \ref{fig7}(b) to \ref{fig7}(f). However, the calculated CLDs in Figs. \ref{fig8}(b) to \ref{fig8}(f) show a slight mismatch with the experimental data unlike the case of Figs. \ref{fig7}(b) to \ref{fig7}(f) where the match is near perfect. 

A likely reason for the different levels of agreement between calculated data with the two models and experimental data is that different kinds of approximations were made in the formulation of the models. The VSM model considers all possible 3 D orientations of the cylinder in the computation of the cylindrical PDF \cite{Vaccaro2006}. However, the LW model considers only one 2 D projection of the ellipsoid where the major and minor axes are parallel to the $x-y$ plane \cite{Li2005n1}. Also, the cylindrical shape of the VSM model is closer to the needle shape of the particles than the ellipsoidal shape of the LW.

Figure \ref{fig9}(a) shows the volume based PSD calculated with the VSM model plotted as a function of the characteristic length $l_c = a_c$ (the length of the cylinder). This data can be compared to the Feret Max data in Fig. \ref{fig3}(b). The Feret Max data covers a range of about $100\mu$m to about $800\mu$m for samples 4 and 5, and about  $10\mu$m to about $500\mu$m for samples 1 to 3. The characteristic lengths predicted by the VSM model in Fig. \ref{fig9}(a) are short of the Feret Max data in Fig. \ref{fig3}(b) because the aspect ratios predicted by the VSM model in Fig. \ref{fig6}(b) are higher than the estimated aspect ratios in Fig. \ref{fig3}(d). This implies that the VSM model predicts needles that are slightly thicker and shorter than the actual needles in the samples. However, the needle lengths calculated with the VSM model cover a range of about $10\mu$m to about $300\mu$m which are still comparable to the Feret max measurements in Fig. \ref{fig3}(b).

A similar situation holds for the LW model where the predicted ellipsoid heights ($l_e$ in Fig. \ref{fig9}(b)) are short of the Feret Max measurements in Fig. \ref{fig3}(b). Similarly, the aspect ratios predicted by the LW model in Fig. \ref{fig6}(d) are higher than the estimated aspect ratios in Fig. \ref{fig3}(d). This again shows that the LW model predicts needles which are slightly thicker and shorter than the actual needles in the samples. The range of needle lengths calculated with the LW model are reasonable when compared with the measured Feret Max in Fig. \ref{fig3}(b).

Even though the predicted lengths ($l_c$ and $l_e$) do not have a perfect match with the  measured Feret Max data, the trend in the lengths of needles from sample 1 to sample 5 in Fig. \ref{fig3}(b) are consistent with the trend in needle lengths from sample 1 to sample 5 in Fig. \ref{fig9}(a). However, the trend in needle lengths in Fig. \ref{fig9}(b) are not so consistent with the trend in needle lengths in Fig. \ref{fig3}(b) moving from sample 1 to sample 5. This is because the LW model predicts smaller aspect ratios for sample 2 and sample 3 in Fig. \ref{fig6}(d) resulting in a shift of the distributions to higher values for sample 2 and sample 3 in Fig. \ref{fig9}(b).

\section{Conclusions}
\label{disc}


We have presented an algorithm which produces best estimates of PSD and particle aspect ratio from measured CLD data. Although the algorithm does not require any additional information about particle size range or particle aspect ratio, this information can be used to further constrain the search if it is available.  If such information is not available (for example during in situ monitoring of a crystallisation process), then the algorithm will perform an automatic search for the best estimate of particle size range and aspect ratio. The approach described here can be used with any geometrical or optical model that provides CLD for particles of given size, shape and optical properties. In the case considered here the particles were treated as opaque and assumed to have convex shapes (that is cylindrical or ellipsoidal). This representation is suitable for the CoA particles considered here as can be seen in Fig. \ref{fig1}. A more detailed discussion of the possible errors that can occur from using this representation is presented in Section 8 of the supplementary information. Also in the supplementary information is a detailed analysis of sensitivity of resulting estimates to choice of algorithm parameters to validate accuracy and robustness of algorithm outcomes.

We applied the algorithm to previously collected CLD data for slurries of needle shaped crystalline particles of COA with different particle size distributions. COA slurries were characterised using FBRM (to measure CLD), imaging (to measure EQPC, maximum and minimum Feret diameters) and laser diffraction (to measure PSD based on equivalent sphere diameter approximation). Measured CLD data were used in the algorithm without any further information input, using two different CLD geometrical models, one for ellipsoids and the other one for thin cylinders. Best estimates for particle aspect ratios and corresponding PSDs were obtained with each model and these were compared to experimental data from imaging and laser diffraction.

Estimated aspect ratios from the thin cylinder model were in good agreement with those obtained from the ratio of maximum and minimum Feret diameters, while those from the ellipsoid model were somewhat higher. Corresponding to this, there was a good agreement between measured and fitted CLDs for the thin cylinder model, but some discrepancies could be seen for the ellipsoid model. Ranges and modes of particle size distributions determined for both models were in a good agreement with those obtained by imaging.
Although it was possible to estimate aspect ratios of needle like particles from CLD data reasonably accurately for the system analysed here, the optimisation problem of finding most appropriate PSD and aspect ratio would be greatly simplified if additional information about particle size range or shape is available, for example from a suitable imaging or scattering technique, especially in the case of systems with significant polydispersity or multimodality in terms of particle shape or size.

The technique described here will be of particular benefit to crystallisation process control, since controlling the process critically depends on real-time in situ information about the size and shape of the particulate product.

\section*{Acknowledgement}

The authors wish to thank the EPSRC (grant number EP/K014250/1), AstraZeneca and GlaxoSmithKline for generous funding for this project.

\newpage
\setcounter{equation}{0}
\setcounter{section}{0}
\setcounter{figure}{0}

\begin{center}
\section*{\huge{Supplementary Information}}
\end{center}

\section{Slack Variables}

The concept of slack variables in optimisation problems is described in previous literature \cite{Boyd2004}. The idea of introducing columns of 1s to the transformation matrix is based on the following argument. Consider the optimisation problem: \\
find $\beta$ which minimises the objective function $\phi$ where
      \begin{equation}
      \phi = \sum_{i=1}^M{\left[y_i - g_i(\beta)\right]^2},
      \label{eqs1}
      \end{equation}
   where $y\in \mathbb{R}^M$, $\beta\in\mathbb{R}^N$ and $g: \mathbb{R}^N\rightarrow\mathbb{R}$. The optimisation problem in Eq. \eqref{eqs1} is equivalent to
         \begin{equation}
         \begin{split}
         \textrm{minimise} &~~ \sum_{i=1}^M{z_i} \\
         \textrm{subject to} &\quad  z_i = \left[y_i - g_i(\beta)\right]^2.
         \label{eqs2}
         \end{split}
         \end{equation}
  Since $[y_i - g_i(\beta)]^2 \geq 0$, then $z_i \geq y_i - g_i(\beta)$. Hence the optimisation problem in Eq. \eqref{eqs2} is equivalent to  
         \begin{equation}
         \begin{split}
         \textrm{minimise} &~~\sum_{i=1}^M{z_i} \\
         \textrm{subject to} & \quad y_i - g_i(\beta) -  z_i \leq 0.
         \end{split}
         \label{eqs3}
         \end{equation}
   There exist slack variables $s_i\geq 0,~j=1,2, \ldots, M$ such that $y_i - g_i(\beta) -  z_i + s_i = 0$. Hence the optimisation problem in Eq. \eqref{eqs3} is equivalent to
           \begin{equation}
           \begin{split}
           \textrm{minimise} &~~\sum_{i=1}^M{z_i} \\
          \textrm{subject to} & \quad y_i - g_i(\beta) -  z_i + s_i = 0 \\
          					& \quad s_i \geq 0.
          \end{split}
           \label{eqs4}
           \end{equation}
   Substituting for $z_i$ in Eq. \eqref{eqs4} gives the following equivalent formulation for the optimisation problem in Eq. \eqref{eqs1}:
         \begin{equation}
         \begin{split}
         \textrm{minimise} &~~\sum_{i=1}^M{y_i - g_i(\beta) + s_i} \\
         \textrm{subject to} & \quad s_i\geq 0 \quad \square .
         \end{split}
         \label{eqs5}
         \end{equation}
      
\section{Choice of Algorithm Parameters}
\label{param}

  \begin{figure}[tbh]
 \centerline{\includegraphics[width=\textwidth]{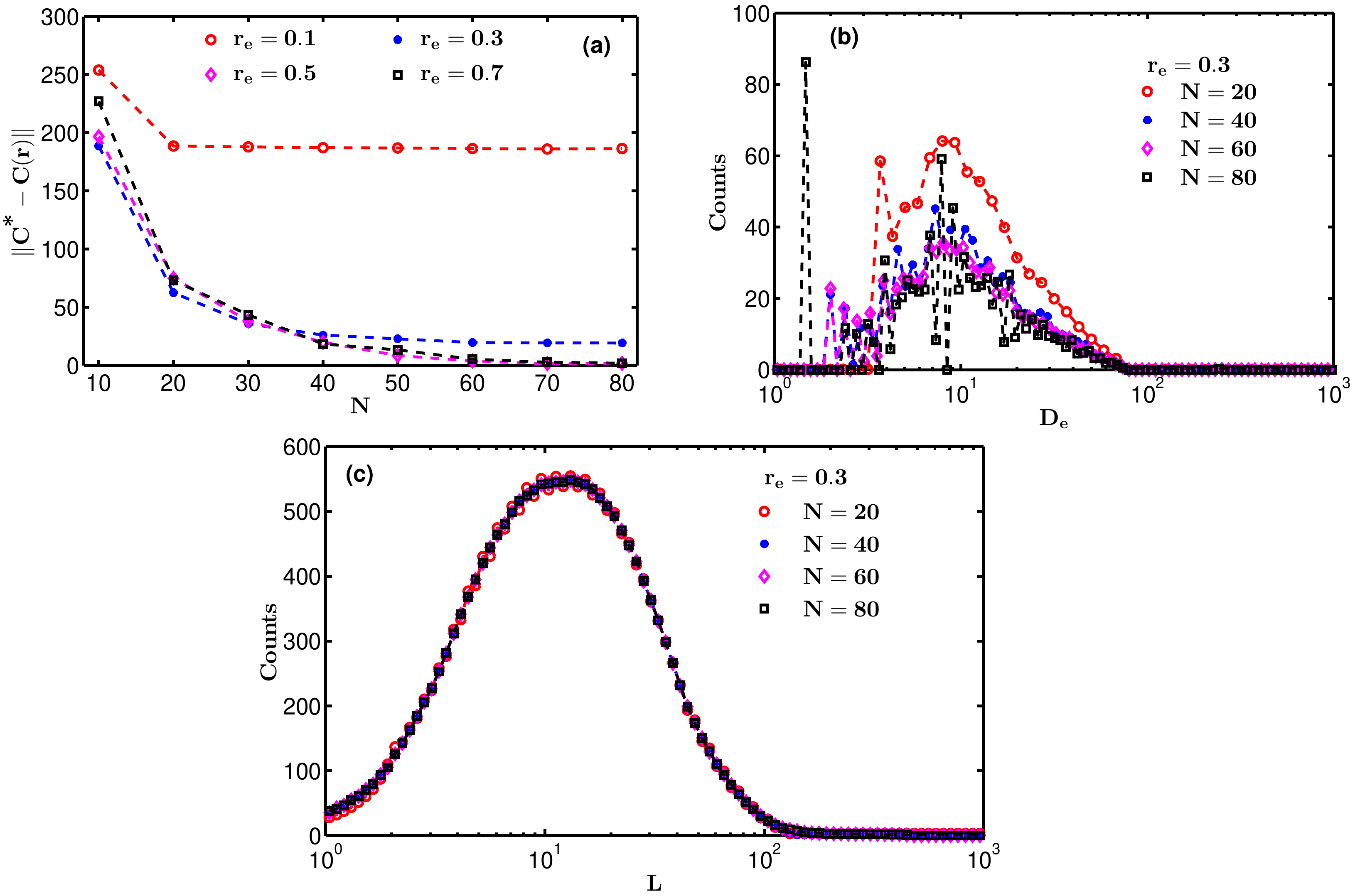}}
 \caption{(a)\,Variation of the $L_2$ norm in Eq. 25 of the main text (from the LW model) with the number of size bins $N$ at the different aspect ratios $r_e$ (indicated in the Figure) for Sample 1. (b)\,Recovered number distributed PSDs (from the LW model) at the specified values of $r_e$ and $N$. The quantity $D_e$ is the characteristic size for the LW model described in the main text. (c)\,Chord length distributions corresponding to the PSDs in (b). The parameter $L$ is the chord length described in the main text.}
 \label{figs1}
  \end{figure}
   \begin{figure}[tbh]
  \centerline{\includegraphics[width=\textwidth]{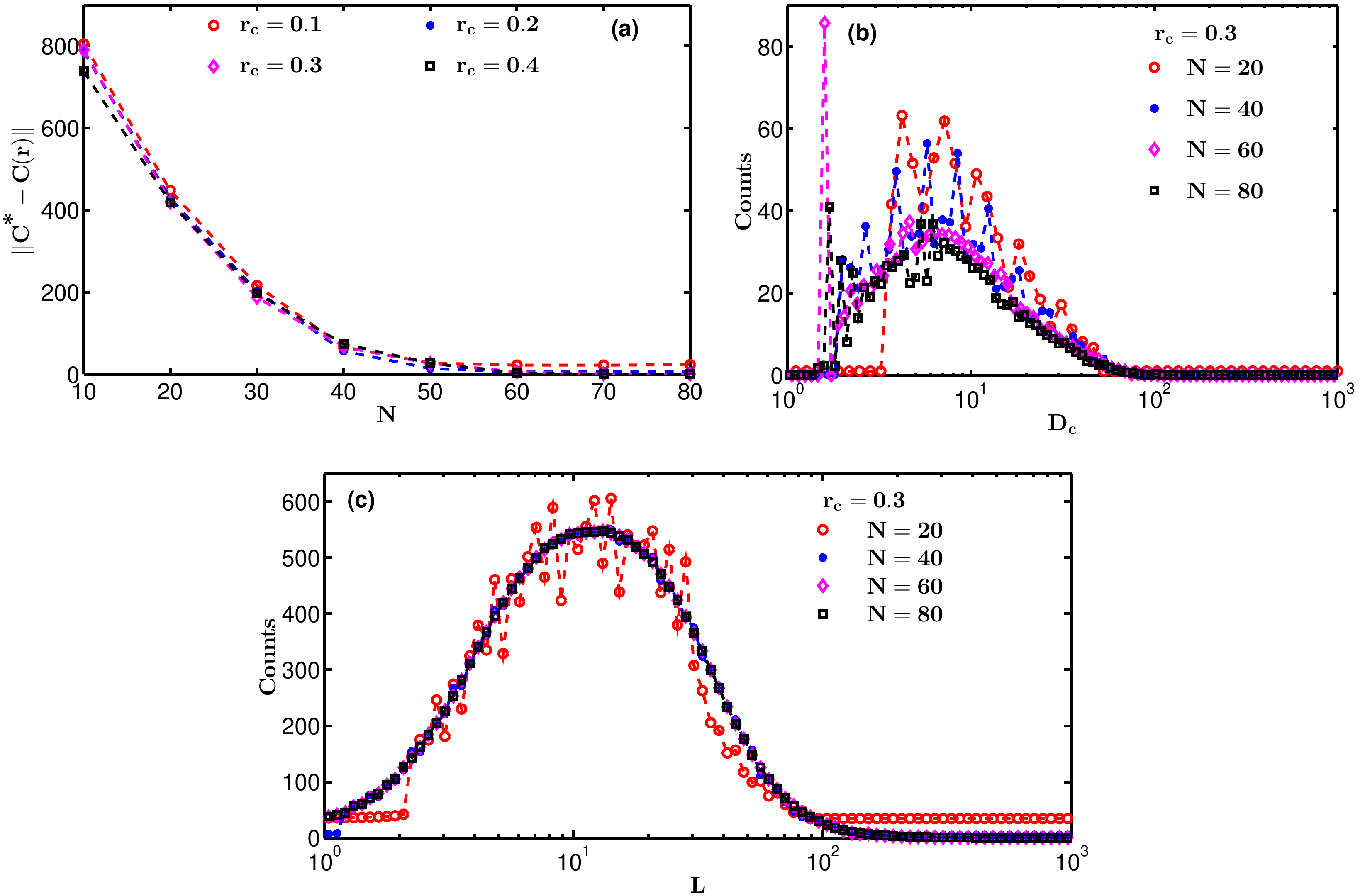}}
  \caption{Similar to Fig. \ref{figs1} obtained with the VSM model.}
  \label{figs2}
   \end{figure}
  
    \begin{figure}[tbh]
   \centerline{\includegraphics[width=\textwidth]{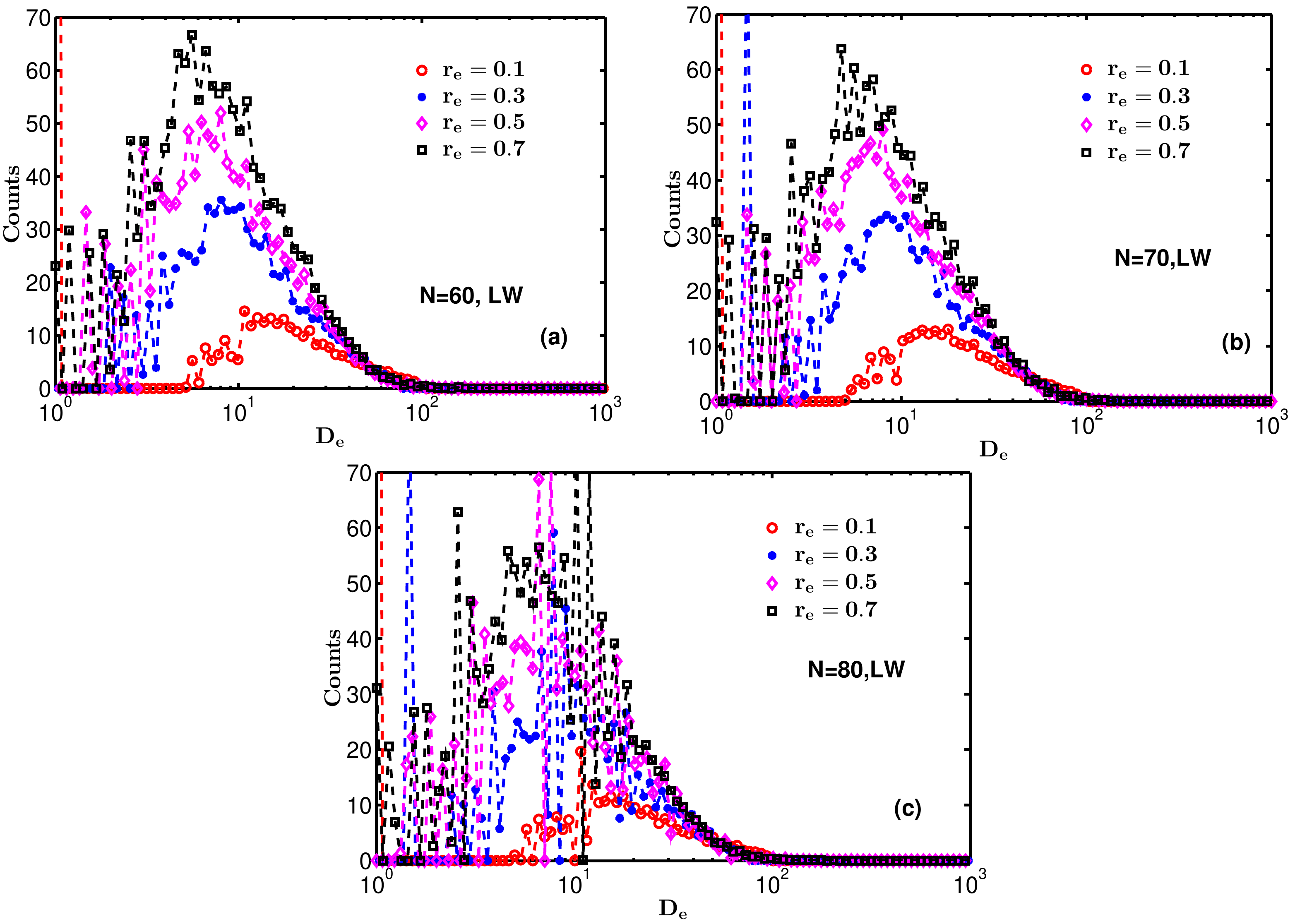}}
   \caption{Particle size distributions recovered (using the LW model at the aspect ratios $r_e$ indicated)  by minimising the objective function $f_1$ in the main text using the different number of particle size bins $N$ indicated in each figure.}
   \label{figs3}
    \end{figure}
    \begin{figure}[tbh]
   \centerline{\includegraphics[width=\textwidth]{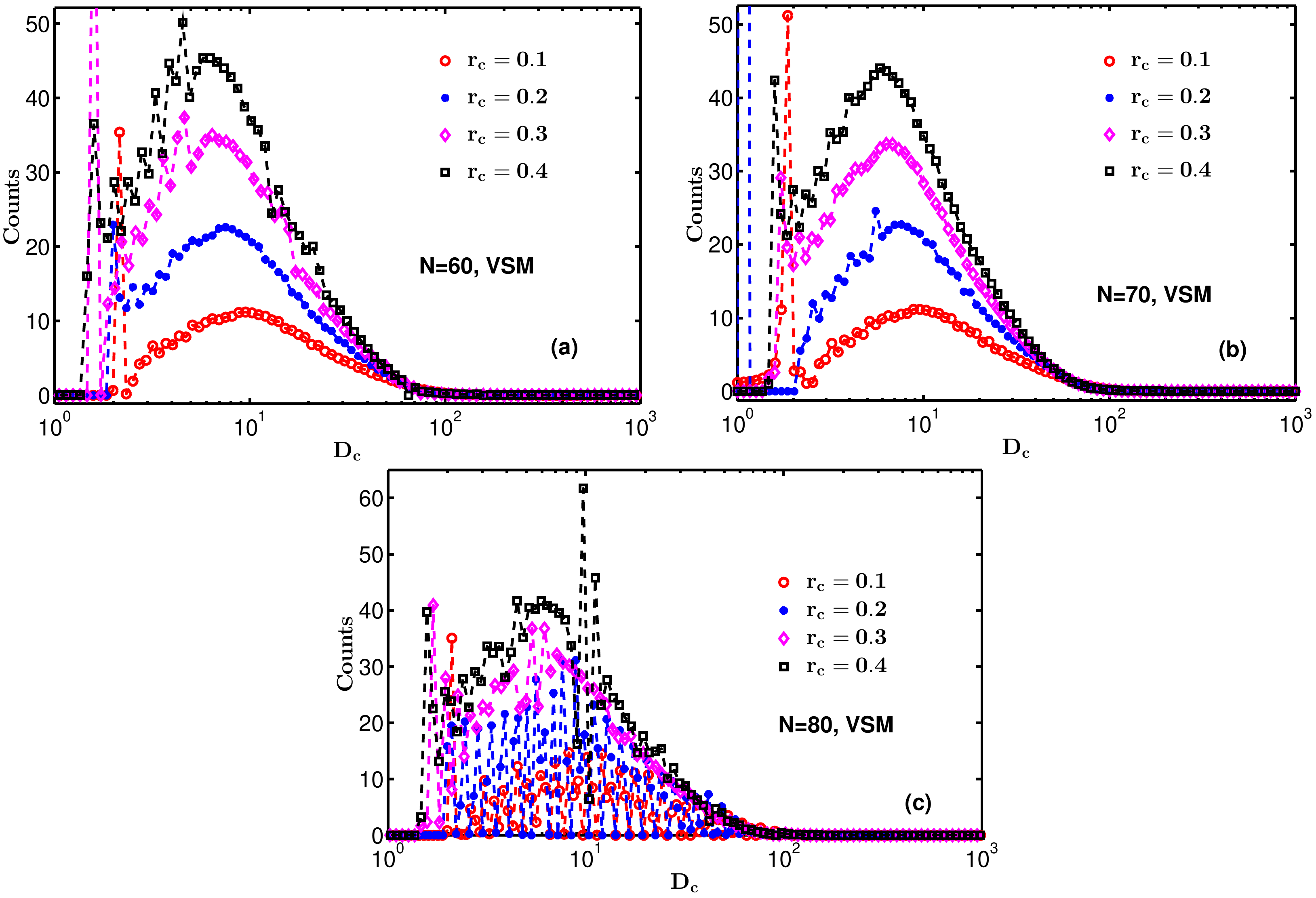}}
   \caption{Same as in Fig. \ref{figs3} with the VSM model.}
   \label{figs4}
    \end{figure}
   
    \begin{figure}[tbh]
   \centerline{\includegraphics[width=\textwidth]{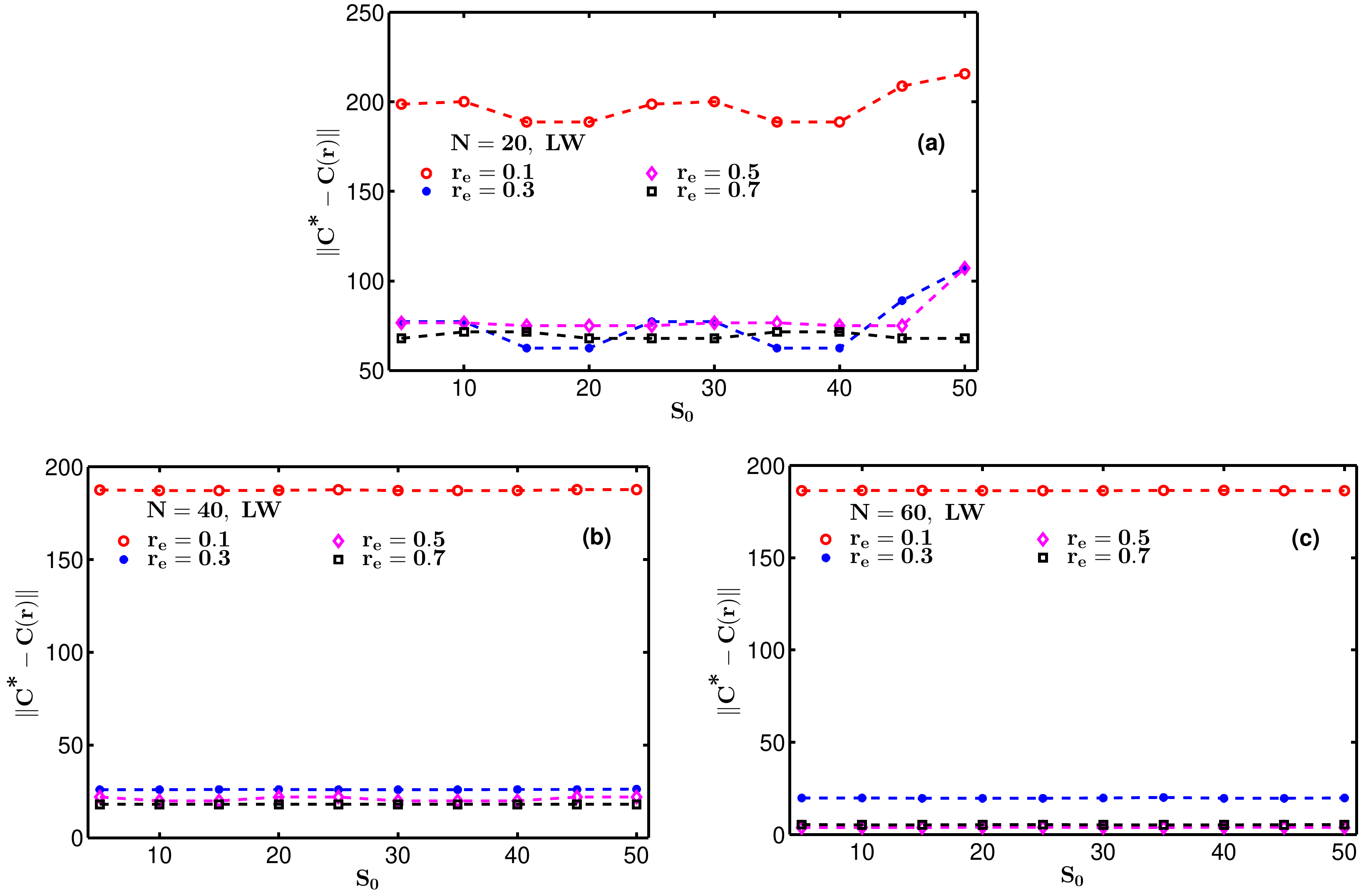}}
   \caption{Variation of the $L_2$ norm in Eq. 25 of the main text with different initial values of window size $S_w$. The calculations were done with the LW model at the different aspect ratios $r_e$ and number of particle size bins $N$ indicated in each figure.}
   \label{figs5}
    \end{figure}
   
     \begin{figure}[tbh]
    \centerline{\includegraphics[width=\textwidth]{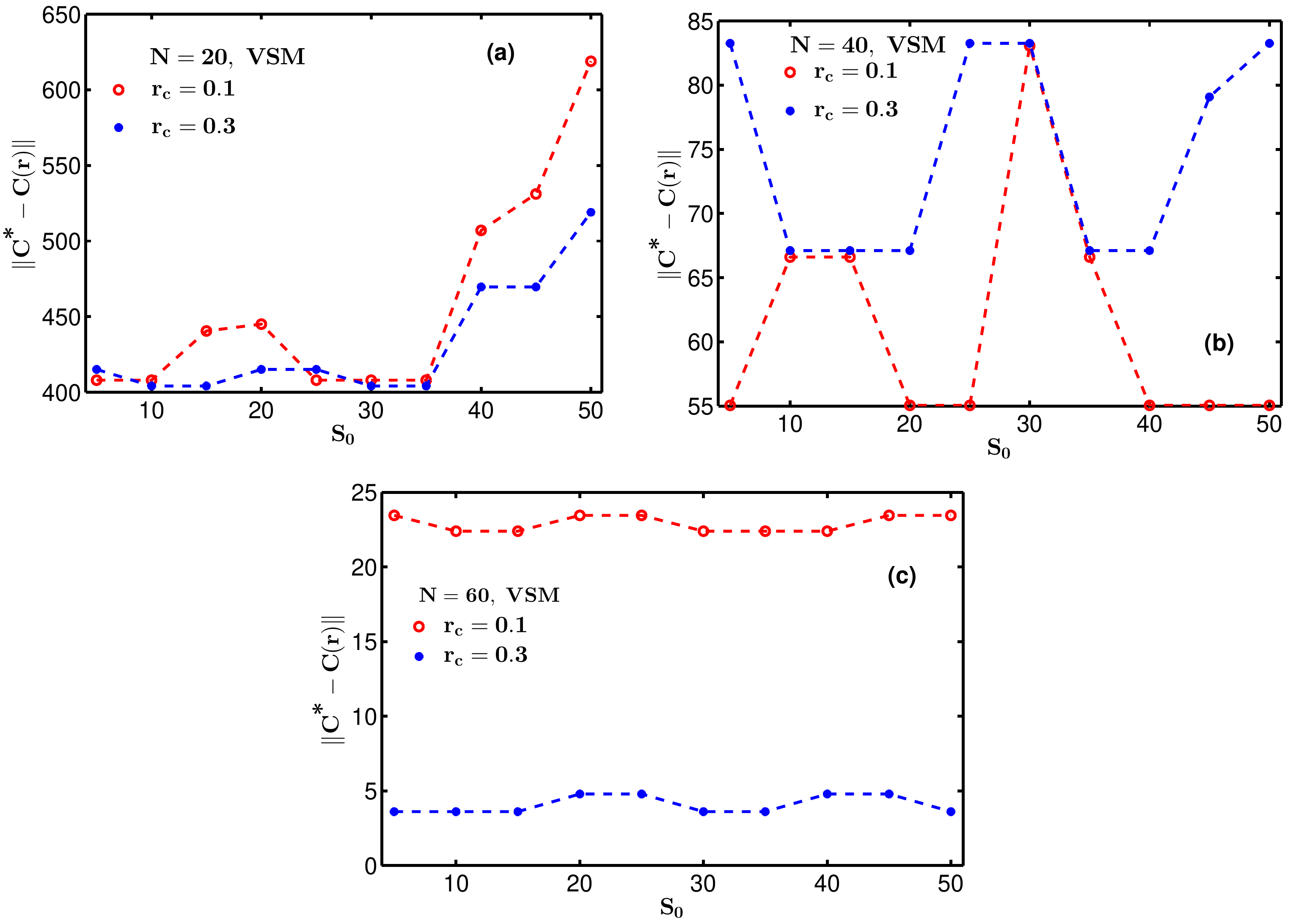}}
    \caption{Same as in Fig. \ref{figs5} with the VSM model.}
    \label{figs6}
     \end{figure}

In this section the motivations for choice of values for parameters in the inversion algorithm are presented.

\subsection{Number of size bins N}

The solution vector $\mathbf{X}$ which minimises the objective function $f_1$ in Eq. (24) of the main text varies slightly with different numbers of particle size bins $N$. This in turn leads to a variation in the vector $\mathbf{C}$ obtained from the forward problem in Eq. (4) of the main text. Hence different values of $N$ were used and each time the $L_2$ norm in Eq. (25) of the main text was calculated in order to determine the optimum number of fitting parameters. 

The variation of the $L_2$ norm with the number of particle size bins $N$ at different aspect ratios for the LW model is shown in Fig. \ref{figs1}(a). As the value of $N$ increases, the $L_2$ norm decreases gradually and then begins to level off at large values of $N$. The result is the same for different aspect ratios $r_e$ as in Fig. \ref{figs1}(a). For a fixed aspect ratio $r_e$ (for example $r_e = 0.3$ in Fig. \ref{figs1}(b)) and a small value of $N$, the PSD obtained from the inverse problem is a bit noisy at the left hand tail of the distribution as in the case of $N=20$ in Fig. \ref{figs1}(b), while the corresponding CLD calculated from the forward problem contains small oscillations as shown in Fig. \ref{figs1}(c). As the value of $N$ is increased, the recovered PSD becomes more noisy as can be seen for the case of $N=40$ in Fig. \ref{figs1}(b). However, the oscillations in the corresponding CLD decrease as in Fig. \ref{figs1}(c). As $N$ is increased further, the oscillations in the recovered PSD become more severe as in Fig. \ref{figs1}(b) for $N=80$. The corresponding CLD for $N=80$ shows very little change from that obtained at $N=40$. 

A similar situation holds for the VSM model where the $L_2$ norm levels off with increasing $N$ as in Fig. \ref{figs2}(a) for different aspect ratios $r_c$. The behaviour of the recovered PSDs for different values of $N$ in Fig. \ref{figs2}(b) is similar to the case of Fig. \ref{figs1}(b). Also, the behaviour of the corresponding CLDs for different values of $N$ in Fig. \ref{figs2}(c) is similar to the case of Fig. \ref{figs1}(c). 

Figures \ref{figs1}(a) and \ref{figs2}(a) show that the $L_2$ norm had become fairly level for $N\gtrsim 60$ for both models and all aspect ratios, which suggests that the calculations reach about the same level of accuracy for number of particle size bins  $N\gtrsim 60$. However, as already seen in Fig. \ref{figs1}(b) and \ref{figs2}(b) the recovered PSDs have different levels of fluctuations for $N\gtrsim 60$. This situation is shown more clearly in Figs. \ref{figs3} and \ref{figs4}.

Figure \ref{figs3}(a) shows the recovered PSD (with the LW model) at the indicated aspect ratios $r_e$ for $N=60$. The PSD for $r_e = 0.1$ is fairly smooth except the long spike at $D_e\approx 1$. However, the PSDs begin to develop oscillations as the aspect ratio $r_e$ increases as seen in the cases of $r_e=[0.3, 0.5, 0.7]$ in Fig. \ref{figs3}(a). A similar situation holds for $N=70$ (Fig. \ref{figs3}(b)) and $N=80$ (Fig. \ref{figs3}(c)). However, the oscillations for the case of $N=80$ is much more severe.

Figure \ref{figs4} is similar to Fig. \ref{figs3} but calculated with the VSM model. For a fixed N, the fluctuations in the PSDs increase as the aspect ratio $r_c$ increases as seen in Figs. \ref{figs4}(a), \ref{figs4}(b) and \ref{figs4}(c). The level of fluctuations at $N=80$ in Fig. \ref{figs4}(c) is much more severe when compared with the cases of $N=60$ (Fig. \ref{figs4}(a)) and $N=70$ (Fig. \ref{figs4}(b)).  For $N=60$ (Fig. \ref{figs4}(a)) the small particle sizes of $D_c\approx 2$ for $r_c>0.1$ are not fully resolved when compared with the case of $N=70$ in Fig. \ref{figs4}(b). The data in Figs. \ref{figs1} to \ref{figs4} suggest that the optimum number of size bins $N$ should be $N=70$. This is because the level of accuracy in the calculations does not increase significantly for $N>70$. Instead, using a larger value of $N$ only leads to severe fluctuations in the calculated PSDs and longer computational times. The value of $N=70$ also gives a better resolution of small particle sizes for both models. Hence a value of $N=70$ was used in all the calculations in the main text.

\subsection{Window size S and spacing q}

The inversion algorithm described in Section 4 of the main text places a window of size $S$ on the bins of the chord length histogram. This window starts with an initial size $S_0$, then slides along the bins of the chord length histogram until it reaches the last bin of the chord length histogram. The window then returns to the beginning of the vector at which its size is increased. The calculations are more accurate if the initial window size is sufficiently small. However, this also depends on the number of particle size bins in the particle size histogram. Then the question is: what is the appropriate number of size bins at which the accuracy of the calculations become independent of the initial window size?

Figure \ref{figs5}(a) shows that for $N=20$ (calculations with the LW model), the $L_2$ norm in Eq. 25 of the main text (calculated at the optimum window size and position) shows a dependence on $S_0$ at different aspect ratios. This dependence reduces significantly at $N=40$ as in Fig. \ref{figs5}(b) and becomes nearly independent at $N=60$. 

A similar situation holds for calculations with the VSM model where the large dependence of the $L_2$ norm (at different aspect ratios) on $S_0$ seen in Fig. \ref{figs6}(a) (for $N=20$) decreases as $N$ increases to 40 in Fig. \ref{figs6}(b). The $L_2$ norm becomes nearly independent of $S_0$ at $N=60$ as in Fig. \ref{figs6}(c). The values of the $L_2$ norm obtained with the VSM model for $N\gtrsim 40$ (Fig. \ref{figs6}) are significantly less than the values of the $L_2$ norm obtained with the LW model for the same aspect ratios in Fig. \ref{figs5}. This suggests that the cylindrical geometry of the VSM model fits the needle data better than the ellipsoidal geometry of the LW for sufficiently large $N$.

The results in Figs. \ref{figs5} and \ref{figs6} suggest that any value of $S_0$ from 2 up to 50 (corresponding to a particle size range of about $1\mu$m to about $43\mu$m) could be used in the calculations for $N\gtrsim 60$. However, a value of $S_0 = 6$ (corresponding to a particle size range of $1\mu$m to $1.5\mu$m) and $N=70$ were used in all the calculations in the main text. The spacing between consecutive positions (that is, $q$ in Eq. 17 of the main text) was kept at $q=2$ in all the calculations in the main text. The smallest value of $q=1$ did not yield any significant increase in accuracy of the calculations.

\section{Length Weighting}
\label{lenwt}

\begin{figure}[tbh]
  \centerline{\includegraphics[width=\textwidth]{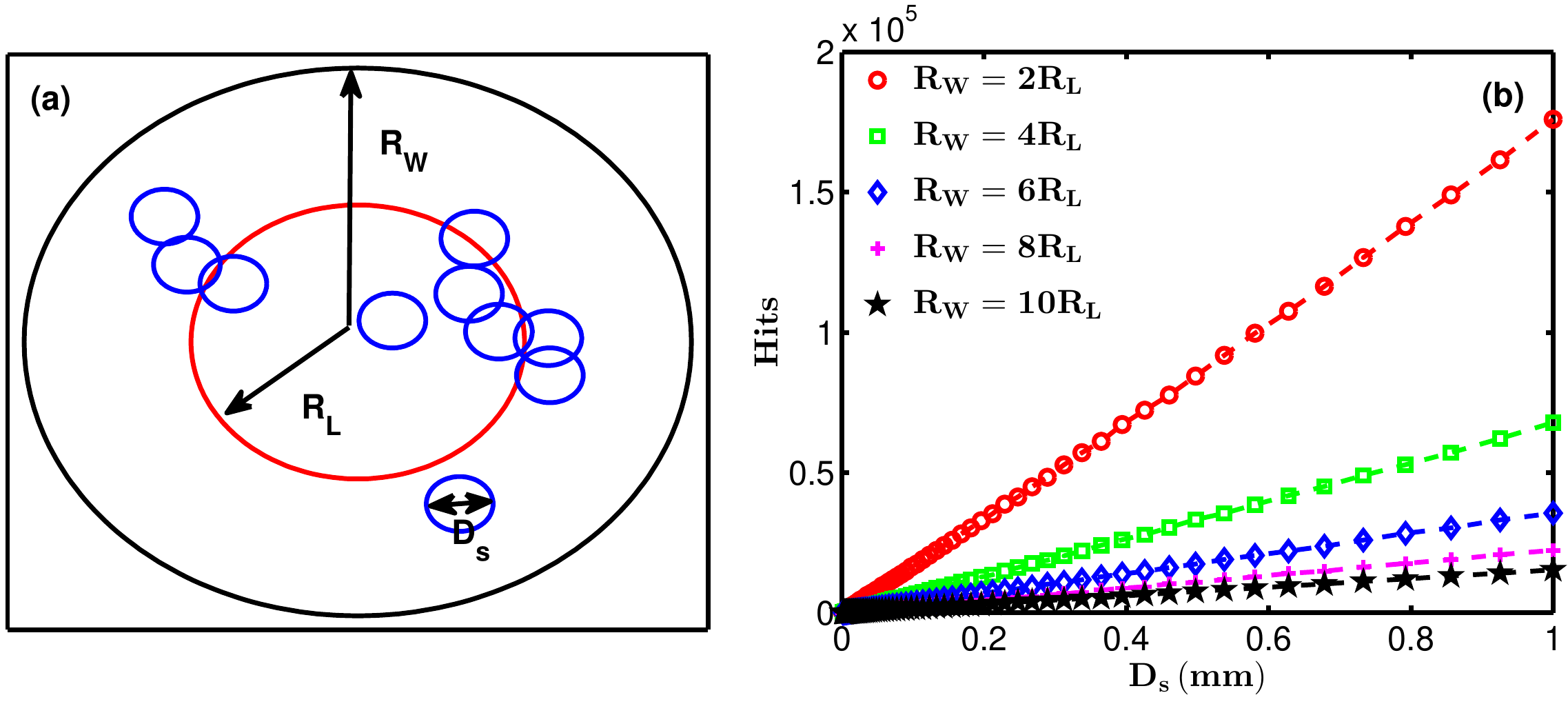}}
  \caption{(a)\,Pictorial representation of the viewing window (black circle) of the FBRM probe, the laser beam (in the focal plane) is represented by the red circle while spherical particles are represented by the blue circles. (b)\,Variation of the frequencies of hits of the laser beam with different particles of sizes $D_s$ and different sizes of the viewing window as indicated in the Figure.}
  \label{figs7}
   \end{figure}

In this section we present a simple numerical simulation which demonstrates the effect of particle size on detection probability. It had already been suggested \cite{Hobbel1991,Simmons1999,Vaccaro2006} that larger particles have a higher probability of being encountered by the FBRM laser. Here we represent the laser beam in the focal plane by the red circle in Fig. \ref{figs7}(a). The circular window of the probe is represented by the black circle in Fig. \ref{figs7}(a). We simulate spherical particles (represented by the blue circles in Fig. \ref{figs7}(a)) falling at random positions on the plane of the laser spot. We assume that all particles regardless of size have equal probability of falling in the focal plane. Each time the boundary of a particle intersects the trajectory of the laser beam a `hit' is recorded. The idea behind the simulation is to see how the number of hits scales with the particle size (diameter of each circle).

    \begin{figure}[tbh]
   \centerline{\includegraphics[width=\textwidth]{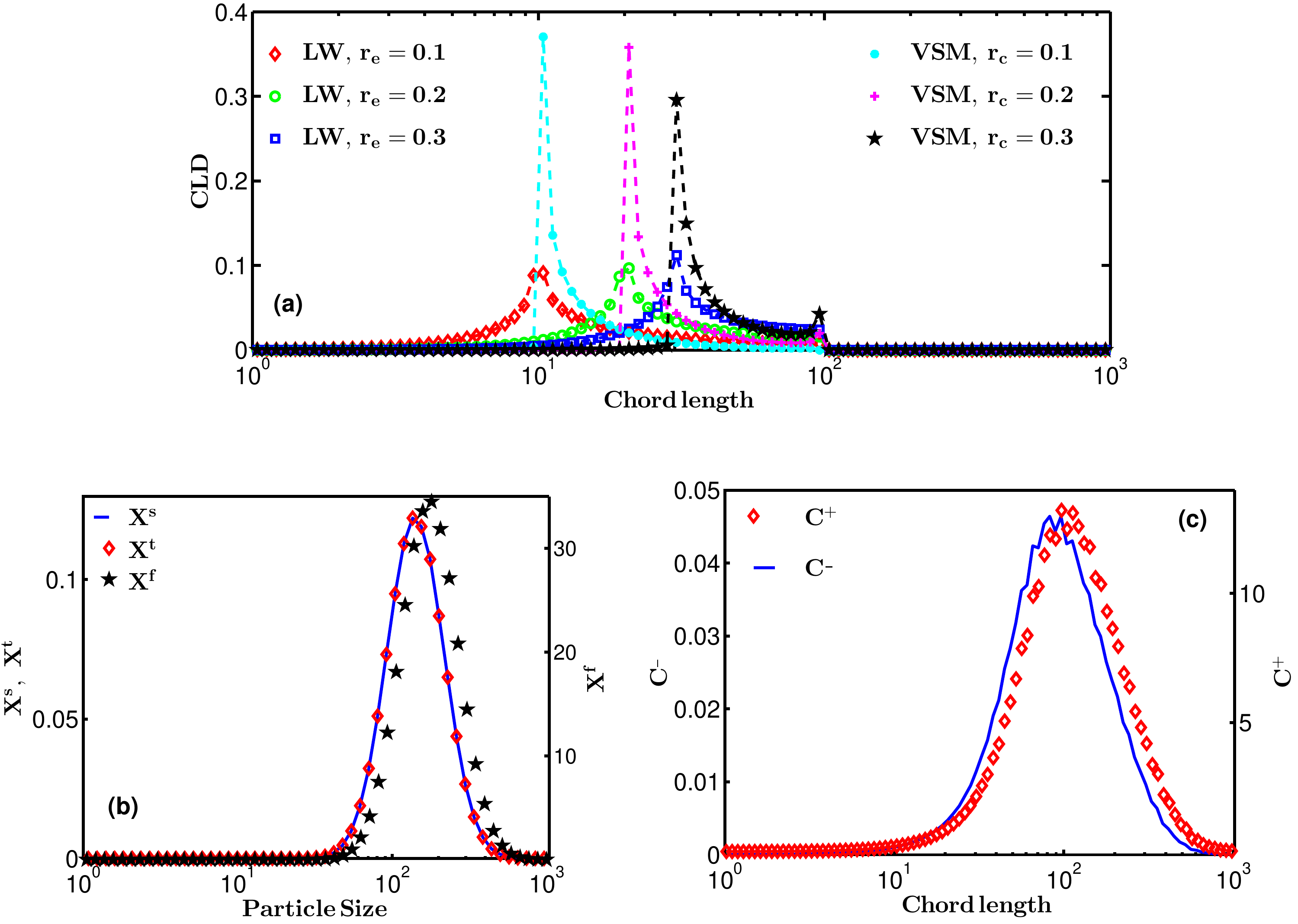}}
   \caption{(a)\,Single particle CLDs for an ellipsoid (for the LW model with aspect ratios indicated as $r_e$) of length $l_e = 2a_e = 100\mu$m ($a_e = $ semi major axis length of ellipsoid) and a cylinder (for the VSM model with aspect ratios indicated as $r_c$) of height $a_c = 100\mu$m. (b)\,Simulated PSD $\mathbf{X}^s$, recovered PSDs $\mathbf{X}^t$ (with length weighted transformation matrix) and $\mathbf{X}^f$ (with unweighted transformation matrix). (c)\,Weighted CLD $\mathbf{C}^+$ from $\mathbf{X}^s$ due to weighted transformation matrix and unweigthed CLD $\mathbf{C}^{\_}$ from $\mathbf{X}^s$ due to unweighted transformation matrix.}
   \label{figs8}
    \end{figure}

Since each event of a particle falling on the focal plane is independent of another particle falling on the focal plane, then we simulate $N_r$ realisations of a single particle of size $D_s$ falling on the focal plane separately from the same number of realisations of another particle of a different size. 

The FBRM probe reports chord lengths between $1\mu$m and $1000\mu$m (for example Fig. 2(b) of the main text). Hence we set the particles sizes $D_s \in [10^{-3}, 1]$mm. The radius $R_L$ of the laser beam is set at 4\,mm \cite{Heinrich2012}, while the radius of the circular window $R_W$ is set in multiples of $R_L$.

The results shown in Fig. \ref{figs7} (b) show that the number of hits scale linearly with the particle size regardless of the size of the probe window. These results agree with earlier suggestions in \cite{Hobbel1991,Simmons1999,Vaccaro2006}. Hence a linear characteristic size weighting is used in the main text in relating the population CLD to the PSD of the population.

\section{Single Particle and Population CLD}

In this section we show the single particle CLD realised with the LW and VSM models. Then we demonstrate the effect of length weighting on the population CLD. 

\subsection{Single Particle CLD of LW and VSM models}

Different mathematical approximations were made in the formulation of the LW and VSM models \cite{Li2005n1,Vaccaro2006} as already noted in the main text. These different approximations give rise to different CLDs for a single particle of similar geometrical shape. The single particle CLDs (for different aspect ratios) realised for an ellipsoid (an ellipse in 2D) of length $l_e = 2a_e = 100\mu$m ($a_e$ is the length of the semi major axis) is shown in Fig. \ref{figs8}(a). The peaks of the single particle CLDs shift to the left as the aspect ratio $r_e = b_e/a_e$ (where $b_e$ is the semi minor axis length) is decreased. The single particle CLDs of the LW model increase slowly at small chord lengths before reaching their peaks at $2b_e$ and then decrease to zero at $l_e$. They have a right shoulder which gets broader as $r_e$ is decreased. The LW model approximates the single particle CLD of the ellipsoid by considering a single projection of the ellipsoid where the major and minor axes are parallel to the $x-y$ plane. It is not known what the effects of the other orientations of the ellipsoid will have on the single particle CLD as these orientations were not considered.

The single particle CLDs of the cylindrical (for a cylinder of height $a_c = 100\mu$m) VSM model shown in Fig. \ref{figs8}(a) are less sensitive to small chord lengths as they rise very quickly to their peaks at $2b_c$ ($b_c$ is the radius of the cylinder). They then decrease more slowly (in a manner similar to the LW case) to zero at $a_c$. The low sensitivity of the single particle cylindrical VSM CLDs to small chord lengths is due to the small angle approximation \cite{Vaccaro2006} made in the calculation of the probability density function for the cylindrical VSM model. However, the positions of the peaks of the single particle cylindrical VSM match those of the LW for the same aspect ratio as seen in Fig. \ref{figs8}(a).

\subsection{Effect of Length Weighting on Population CLD and Recovered PSD}

The effect of the size of a particle to its detection probability has been demonstrated in section \ref{lenwt}. This length bias could have a substantial effect on the calculations if it is not incorporated in some way. Consider the simulated PSD $\mathbf{X}^s$ shown by the solid line in Fig. \ref{figs8}(b). The PSD was made by randomly drawing $10^6$ particle sizes from the normal distribution with mean size 500~$\mu$m and standard deviation 100~$\mu$m. Then the particle sizes were shifted to ensure non negativity. Finally the PSD was made from a normalised histogram of 30 bins. The solid line in Fig. \ref{figs8}(c) shows the CLD $\mathbf{C}^{\_}$ calculated from the normalised PSD $\mathbf{X}^s$ as
               \begin{equation}
          \mathbf{C}^{\_} = \mathbf{A}\mathbf{X}^s,
               \label{eqs6}
               \end{equation}
where $\mathbf{A}$\footnote{The matrix is calculated with the LW model.} is the transformation matrix in Eq. (6) of the main text without any length weighting. The symbols in Fig. \ref{figs8}(c) show the CLD $\mathbf{C}^+$ calculated from the normalised PSD $\mathbf{X}^s$ as 
                 \begin{equation}
            \mathbf{C}^+ = \mathbf{\tilde{A}}\mathbf{X}^s,
                 \label{eqs7}
                 \end{equation}
where $\mathbf{\tilde{A}}$ is the transformation matrix in Eq. (5) of the main text with length weighting. Figure \ref{figs8}(c) shows that the CLD $\mathbf{C}^+$ calculated with length weighting is substantially  higher than the corresponding CLD $\mathbf{C}^{\_}$ without length weighting and slightly shifted to the right. This shows that the experimentally measured CLD could be substantially biased due to the length weighting effect demonstrated in section \ref{lenwt}. Hence the length weighting effect needs to be incorporated into the calculations to account for this length bias.

The red diamonds in Fig. \ref{figs8}(b) show the PSD obtained by minimising the objective function $\phi$ (similar to the function $f_1$ in Eq. 24 of the main text) given as
 			\begin{equation}
            \phi = \sum_{j=1}^{M}{\left[C_j^+ - \sum_{i=1}^N{\tilde{A}_{ji}X_i^t}\right]^2},
             \label{eqs8}
             \end{equation}
  where $M$ is the number of chord length bins, $N$ is the number of particle size bins and $\mathbf{X}^t$ is the optimum PSD which minimises the objective function. The recovered PSD $\mathbf{X}^t$ matches the original PSD $\mathbf{X}^s$ because the length weighting effect has been incorporated into the matrix $\mathbf{\tilde{A}}$. However, when the objective function is formulated as 
  			\begin{equation}
             \phi = \sum_{j=1}^{M}{\left[C_j^+ - \sum_{i=1}^N{A_{ji}X_i^f}\right]^2},
              \label{eqs9}
              \end{equation}
the optimum PSD $\mathbf{X}^f$ is substantially higher than the original PSD $\mathbf{X}^s$ and slightly shifted to the right as seen in Fig. \ref{figs8}(b). This again demonstrates the need to account for the length bias that comes with the experimentally measured CLD to reduce its effect on the calculated PSD.

\section{Number and Volume Based PSD}

Some particle sizing instruments report the PSD in terms of a volume distribution for example Figs. 2(a), 3(a), 3(b) and 3(c) of the main text. Hence it becomes necessary to calculate a volume based PSD that is comparable to the experimentally measured PSDs. The volume based PSD $\mathbf{X}^v$ can be calculated from \cite{Holdich2002}
       \begin{figure}[tbh]
      \centerline{\includegraphics[width=\textwidth]{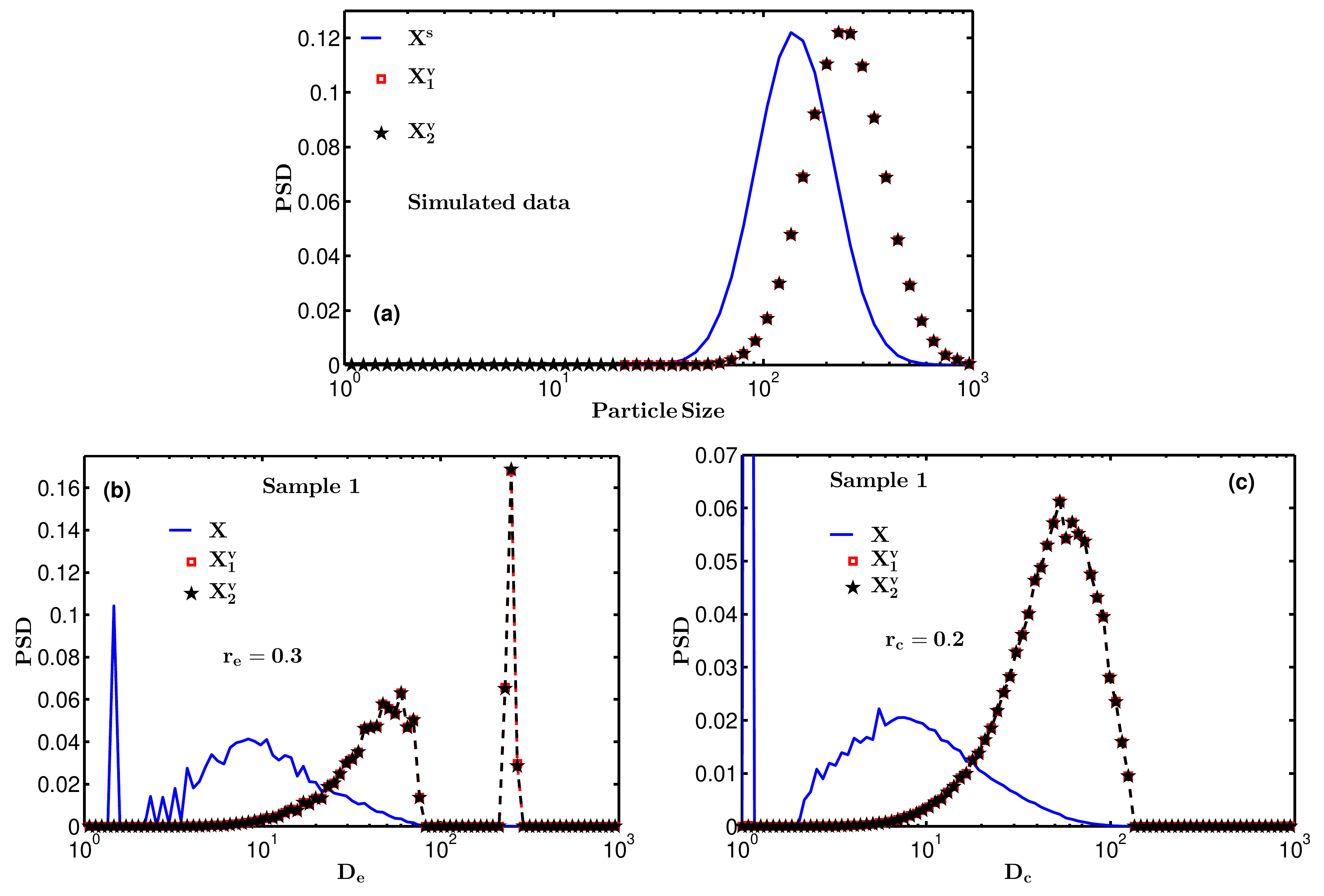}}
      \caption{(a)\,The simulated PSD $\mathbf{X}^s$ in Fig. \ref{figs8}(b), volume based PSDs $\mathbf{X}^v_1$ (calculated from Eq. \eqref{eqs10} using $\mathbf{X}^s$) and $\mathbf{X}^v_2$ (calculated from Eq. \eqref{eqs13}). (b)\,Normalised number based PSD $\mathbf{X}$ obtained by minimising the function $f_1$ in the main text using the LW model at the aspect ratio $r_e$ indicated in the figure. Volume based PSD $\mathbf{X}_1^v$ calculated from Eq. \eqref{eqs10} using the PSD $\mathbf{X}$. Normalised volume based PSD $\mathbf{X}_2^v$ obtained from the function $f_3$ (at $\lambda = 0$) in the main text. (c)\,Same as in (b) with the VSM model.}
      \label{figs9}
       \end{figure}
      
  			\begin{equation}
            X_i^v = \frac{X_i\overline{D}_i^3}{\sum_{i=1}^N{X_i\overline{D}_i^3}},
              \label{eqs10}
              \end{equation}
 where $\mathbf{X}$ is the number based PSD and $\overline{D}$ is the characteristic size of the population of particles. This is equivalent to 
   			\begin{equation}
             X_i^v = \frac{\hat{X}_i\overline{D}_i^3}{\sum_{i=1}^N{\hat{X}_i\overline{D}_i^3}},
               \label{eqs11}
               \end{equation}
    where
   			\begin{equation}
             \hat{X}_i = \frac{X_i}{\sum_{i=1}^N{X_i}}.
               \label{eqs12}
               \end{equation}
Because the inversion problem is ill posed, the calculated PSD $\mathbf{X}$ (which is usually Gaussian like) from the experimentally measured CLD could have small fluctuations at the tails of the distribution. The presence of small fluctuations at the right tail of the number based PSD $\mathbf{X}$ leads to artificial peaks at large particle sizes. For example, the number based PSD (shown in Fig. \ref{figs9}(b)) recovered for sample 1 with the LW model contains a small fluctuation at $D_e\approx 200 \mu$m. This leads to the peak at $D_e\approx 200\mu$m in the volume based PSD $\mathbf{X}_1^v$ calculated from Eq. \eqref{eqs10}. This peak is clearly artificial as the number based PSD $\mathbf{X}$ in Fig. \ref{figs9}(b) shows a near zero particle size count at $D_e\approx 200\mu$m. This problem led to the formulation of a new method for calculating the volume based PSD which allows the application of a suitable regularisation to remove these artificial peaks.  

To demonstrate that the method summarised in Eqs. 29 to 33 of the main text reproduces the correct volume based PSD, consider the simulated PSD $\mathbf{X}^s$ in Fig. \ref{figs9}(a) which is the same normalised PSD $\mathbf{X}^s$ in Fig. \ref{figs8}(b). The red squares in Fig. \ref{figs9}(a) show the volume based PSD $\mathbf{X}_1^v$ calculated froom Eq. \eqref{eqs10} using the PSD $\mathbf{X}^s$.
The black pentagrams in Fig. \ref{figs9}(a) show the normalised volume based PSD $\mathbf{X}_2^v$ calculated by minimising the objective function $\phi$ given as
  			\begin{equation}
             \phi = \sum_{j=1}^{M}{\left[C_j^+ - \sum_{i=1}^N{\overline{A}_{ji}\overline{X}^v_i}\right]^2},
              \label{eqs13}
              \end{equation}
 where
     		\begin{equation}
            \overline{A}_{ji} = \frac{\tilde{A}_{ji}}{\overline{D}_i^3},
             \label{eqs14}
              \end{equation}
  $\mathbf{\tilde{A}}$ is the length weighted transformation matrix in Eq. \eqref{eqs7}, $\mathbf{C}^+$ is the length biased CLD in Eq. \eqref{eqs7} and $\mathbf{\overline{X}}_2^v$ is the optimum PSD which minimises the objective function in Eq. \eqref{eqs13}. The normalised volume based PSD $\mathbf{X}_2^v$ obtained as
       		\begin{equation}
             X_{2i}^v = \frac{\overline{X}_{2i}^v}{\sum_{i=1}^N{\overline{X}_{2i}^v}}
               \label{eqs16}
                \end{equation}
   matches the volume based PSD $\mathbf{X}_1^v$ calculated from Eq. \eqref{eqs10}. The PSD $\mathbf{X}_2^v$ is shown by the black pentagrams in Fig. \ref{figs9}(a). The peaks of the volume based PSDs $\mathbf{X}_1^v$ and $\mathbf{X}_2^v$ are shifted to the right of the number based PSD $\mathbf{X}^s$ as expected.
   
   Figure \ref{figs9}(b) shows the volume based PSD $\mathbf{X}_2^v$ calculated (using the LW model and normalised as in Eq. \eqref{eqs16}) by minimising the objective function $f_3$ given by Eq. 33 of the main text for $\lambda = 0$. the volume based PSD $\mathbf{X}_2^v$ calculated from $f_3$ matches the volume based PSD $\mathbf{X}_1^v$ calculated from Eq. \eqref{eqs10} as expected as shown in Fig. \ref{figs9}(b). However, the volume based PSD $\mathbf{X}_2^v$ still contains the artificial peak at $D_e\approx 200\mu$m. This peak can be removed using a suitable value of $\lambda$ which enforces the penalty on the norm as given in Eq. 33 of the main text. The recovered number based PSD $\mathbf{X}$ contains fluctuations at small particle sizes $D_e\approx 1$, but these fluctuations have no effect on the volume based PSDs $\mathbf{X}_1^v$ or $\mathbf{X}_2^v$ since the third moment of particle sizes $D_e\approx 1\mu$m is much less than the third moment of particle sizes $D_e\approx 10^2\mu$m.
   
   A similar situation holds for the VSM model as seen in Fig. \ref{figs9}(c). The volume based PSD $\mathbf{X}_1^v$ calculated from Eq. \eqref{eqs10} using the recovered number based PSD $\mathbf{X}$ matches the volume based PSD $\mathbf{X}_2^v$ obtained by minimising Eq. 33 of the main text. However, in this case, there is no artificial peak in either $\mathbf{X}_1^v$ or $\mathbf{X}_2^v$ at large particle sizes since there are no fluctuations in the number based PSD $\mathbf{X}$ (at large particle sizes) in this case.
      
 \section{Uniqueness of Shape Information}
 
\begin{figure}[tbh]
            \centerline{\includegraphics[width=\textwidth]{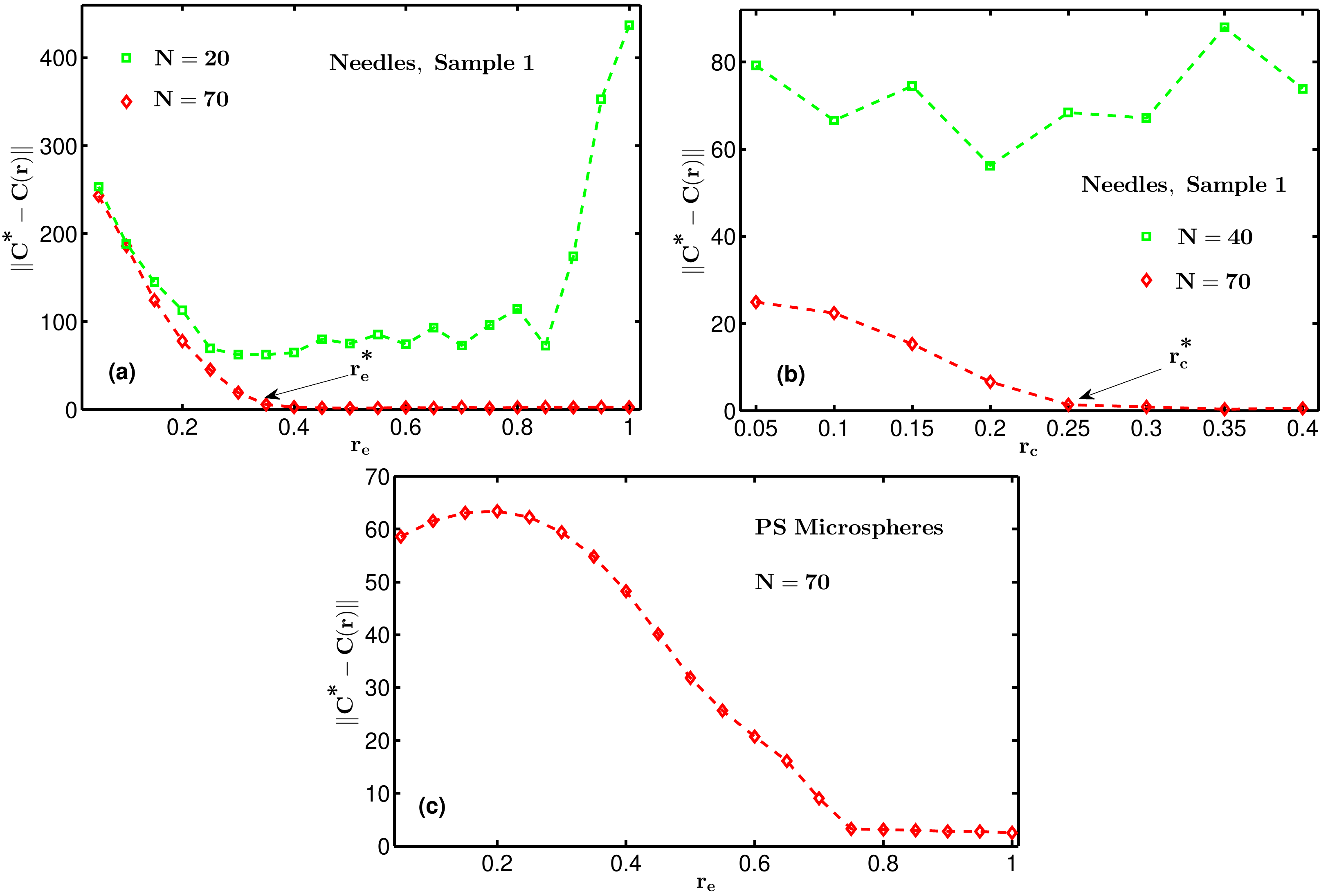}}
            \caption{Variation of the $L_2$ norm in the main text with aspect ratios for different number of particle size bins $N$ for (a)\,sample 1 (calculations with the LW model), (b)sample 1 (calculations with the VSM model) and (c)\,the system of spherical particles described in footnote \ref{fn1} (calculations with the LW model). }
            \label{figs10}
             \end{figure}

  As discussed in section 4 of the main text, minimisation of the objective function $f_1$ defined in Eq. 24 of the main text using different particle size ranges (at different aspect ratios) results in a situation where the $L_2$ norm in Eq. 25 of the main text flattens out after some critical aspect ratio. This situation is shown in Fig. \ref{figs10}(a) (for $N=70$) for sample 1 (referred to in the main text) using the LW model. The situation is the same for the VSM model as seen in Fig. \ref{figs10}(b) for the same sample 1.
 
 The $L_2$ norm decreases with increasing aspect ratio after which it becomes nearly flat after some critical aspect ratio indicated as $r_e^{*}\approx 0.3$ in Fig. \ref{figs10}(a) and $r_c^{*}\approx 0.25$ in Fig. \ref{figs10}(b). The $L_2$ norm in the two cases shown in Figs. \ref{figs10}(a) and \ref{figs10}(b) does not have a clear global minimum making it necessary to reformulate the problem in such a way that unique shape information can be retrieved. The problem of non uniqueness is common to other samples discussed in the main text. 
 
                       \begin{figure}[tbh]
                      \centerline{\includegraphics[width=\textwidth]{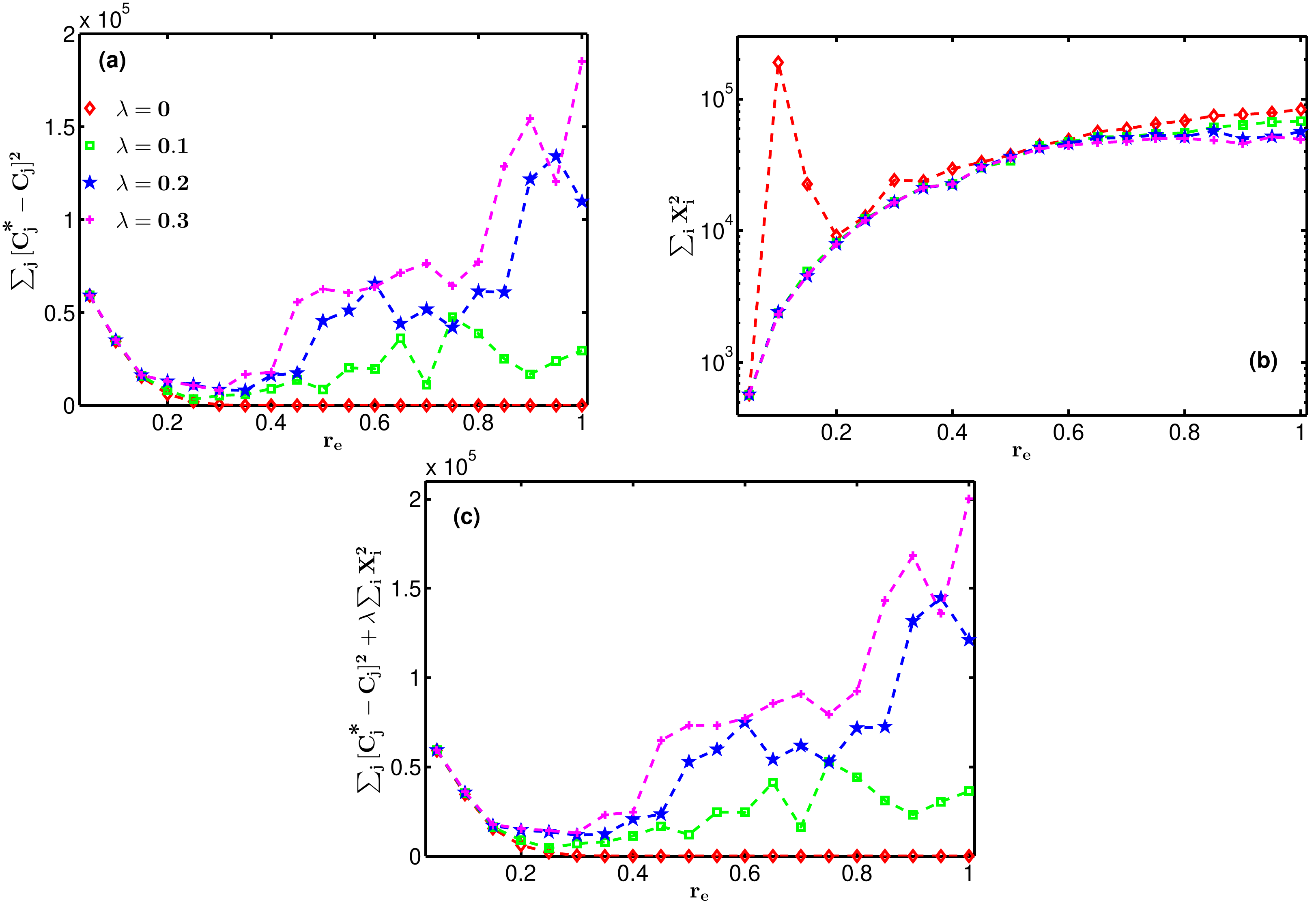}}
                      \caption{(a)\,Variation of the squared residual norm between the measured CLD $\mathbf{C}^{\textbf{*}}$ and calculated CLD $\mathbf{C}$ with aspect ratio for the different values of $\lambda$ (in the function $f_2$ in the main text) indicated in the figure. (b)\,Variation of the square norm of the recovered PSD with aspect ratio for the values of $\lambda$ indicated in (a). (c)\,Variation of the function $f_2$ with aspect ratio for the values of $\lambda$ in (a). All calculations with the LW model for sample 1.}
                      \label{figs11}
                       \end{figure}
                      
                                 \begin{figure}[tbh]
                                \centerline{\includegraphics[width=\textwidth]{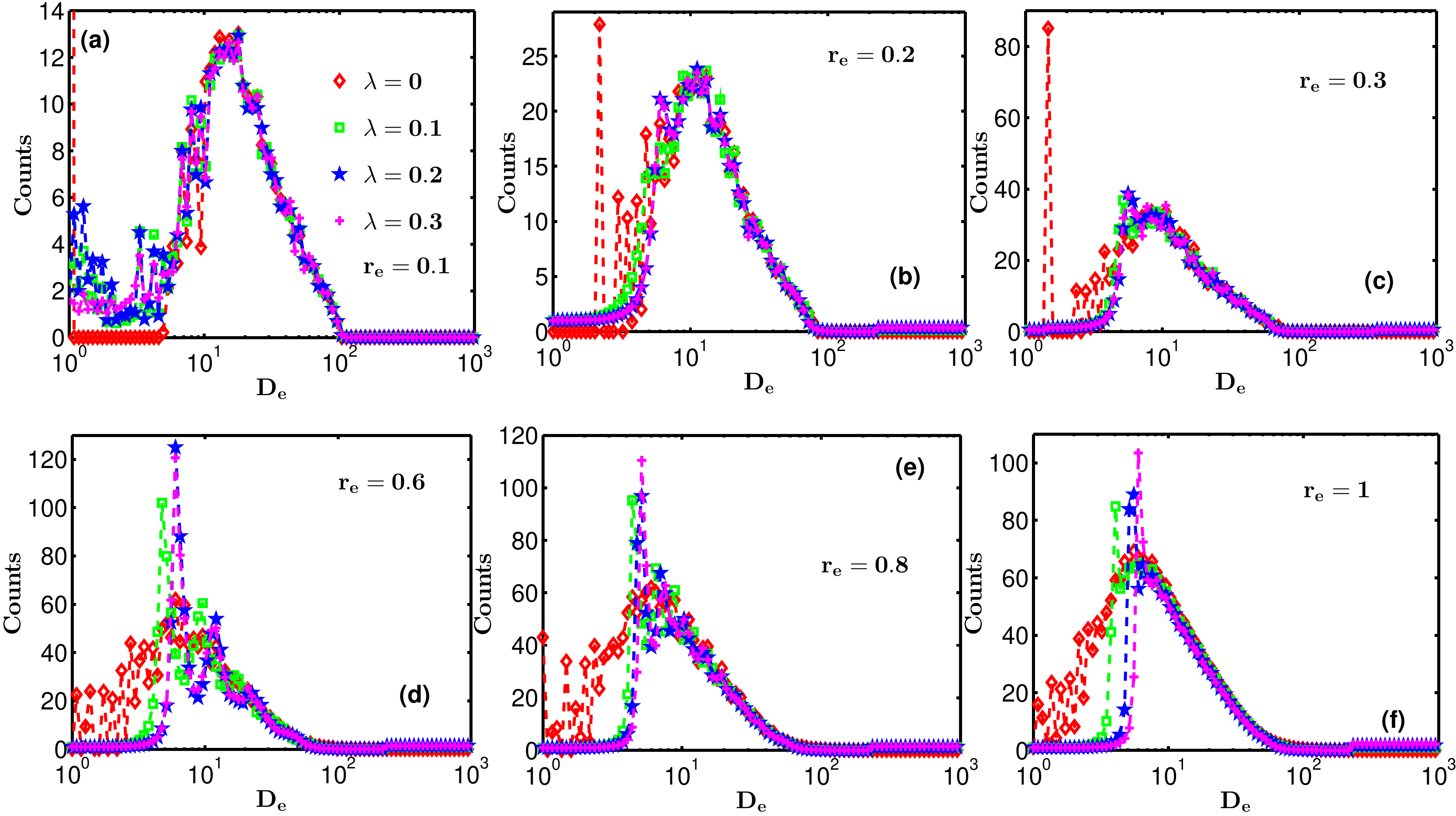}}
                                \caption{The recovered PSDs from the function $f_2$ (at the values of $\lambda$ indicated in (a)) at the aspect ratios $r_e$ indicated in the figures. All calculations with the LW model for sample 1.}
                                \label{figs12}
                                 \end{figure}
                                
                                     \begin{figure}[tbh]
                                      \centerline{\includegraphics[width=\textwidth]{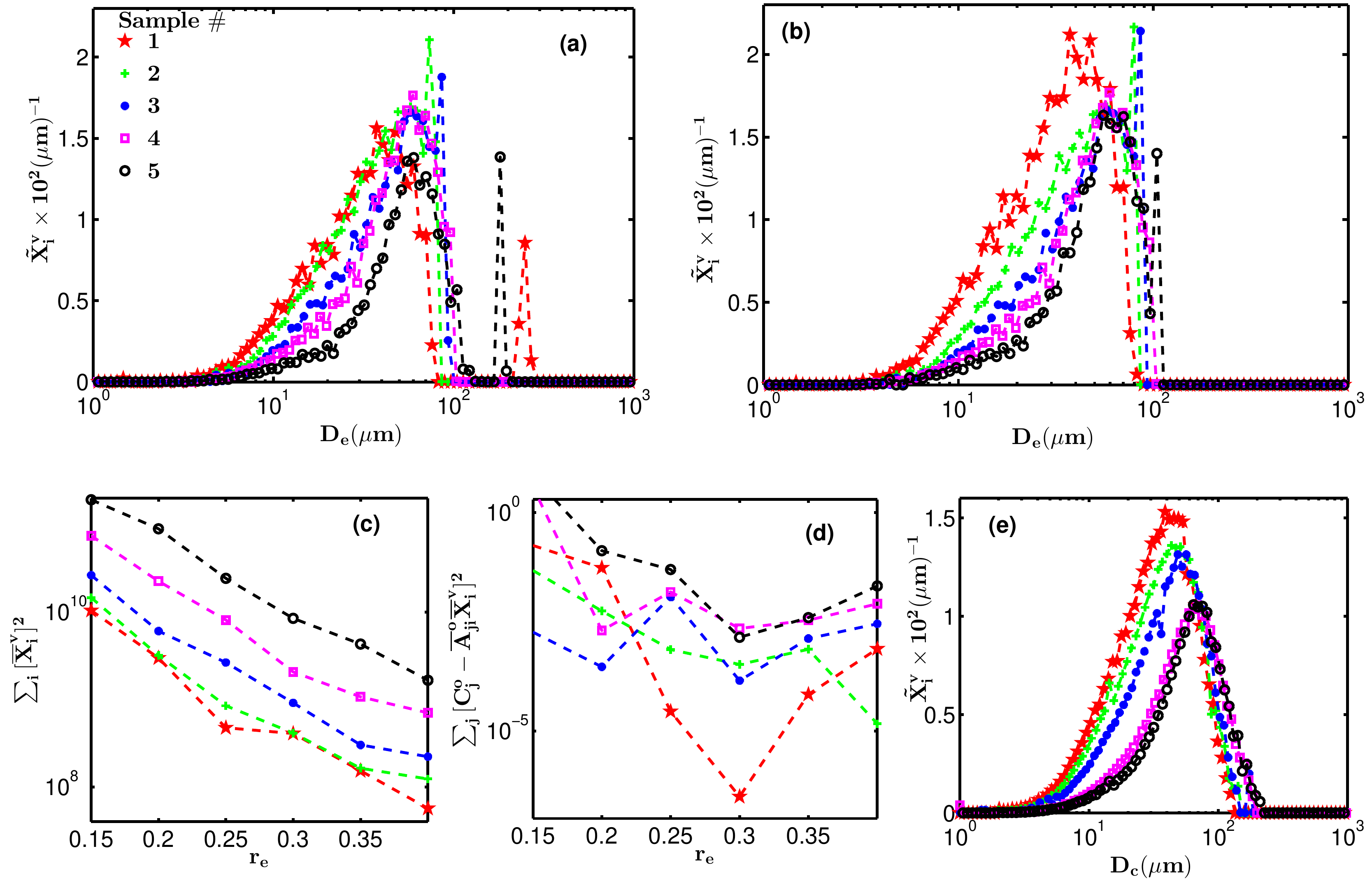}}
                                      \caption{Volume based PSDs calculated from the function $f_3$ at (a)\,$\lambda = 0$ and (b)\,$\lambda = 10^{-14}$ for the five samples in the main text. (c)\,The square norm of the calculated volume based PSD from $f_3$ in the range of aspect ratios where the function $f_2$ reaches it minimum in Fig. \ref{figs11}(c) for all five samples. (d)\,The sum of squared deviation of the calculated CLDs defined in $f_3$ for the range of aspect ratios in (c) for all five samples. All calculations in (a) to (d) done with the LW model. (e)\,Similar to (a) but calculated with the VSM model.}
                                       \label{figs13}
                                        \end{figure}
 This problem of non uniqueness only comes to light when the number of particle size bins is large enough. For the case where the number of particle size bins is not large enough say $N=20$ in Fig. \ref{figs10}(a), an artificial global minimum could be realised for a suitable initial window size. This result is only artificial as it depends on the initial window size chosen. This is because the $L_2$ norm is still dependent on the initial window size as seen in Fig. \ref{figs5}(a). Also the fits obtained at such values of $N$ are poorer than fits obtained at larger $N$ as seen in Fig. \ref{figs5}.
 
 Figure \ref{figs10}(b) also shows a situation where an artificial global minimum is realised with the VSM model for a small value of $N$ ($N=40$ in Fig. \ref{figs10}(b)). The reason being similar to the case of the LW model. The $L_2$ norm is still strongly dependent on the initial window size (for small values of $N$) as seen in Fig. \ref{figs6}.
 
 Figures \ref{figs10}(a) and \ref{figs10}(b) (for $N=70$) show that the level of fit  to the needle data is much better with the VSM than the LW model. This could be because the cylindrical shape of the VSM model is closer to the shape of the needles than the ellipsoid geometry of the LW. This proposition is supported by the fact that for a system of spherical particles\footnote{\label{fn1} The system of spherical particles is a 0.05\% weight suspension of polystyrene microspheres dispersed in isopropanol. The suspension was place in a jacketed vessel (with a jacket temperature of $20^{\circ}$C) in the Mettler Toledo EasyMax\texttrademark~ system. The suspension was stirred at 100rpm and the CLD was measured with a Mettler Toledo FBRM G400 probe.} \footnote{The refernce to the system of spherical particles here is only done for the purpose of comparing Fig. \ref{figs10}(c) with Fig. \ref{figs10}(b). There shall be no further reference to the system of spherical particles beyond this point. All other samples referred to in this text or in the main text are  (or any of) the sample 1 to sample 5 shown in Fig. 1 of the main text.},
 the level of fit obtained with the LW model (Fig. \ref{figs10}(c)) is comparable to the level of fit obtained with the VSM model for the needles (Fig. \ref{figs10}(b)).
 
 This situation of non uniqueness of particle shape information led to the introduction of the objective $f_2$ defined in Eq. 27 of the main text. The motivation comes from the observation in Figs. \ref{figs3} and \ref{figs4} that the level of fluctuations of the recovered number based PSD increases as the aspect ratio increases for a fixed $N$. A possible reason for this could be because the larger aspect ratios in Figs. \ref{figs3} and \ref{figs4} deviate too much from the actual shape of the particles (as seen in Figs. 1 and 3(d) of the main text) even though they yield about the same level of fit with the intermediate aspect ratios as seen in Figs. \ref{figs10}(a) and \ref{figs10}(b) (for $N=70$).
 
 A suitable value of $\lambda$ (in Eq. 27 of the main text) can be chosen by comparing the relative magnitudes of the two sums in Eq. 27 of the main text. The variation of the square residual norm between the measured CLD $\mathbf{C}^{\textbf{*}}$ and calculated CLD $\mathbf{C}$ with aspect ratio for sample 1 (for different values of $\lambda$ in Eq. 27 of the main text) are shown in Fig. \ref{figs11}(a). The square of residual norm for $\lambda = 0$ (in the flat region) is of the order of $10^4$ as seen in Fig. \ref{figs11}(a). Figure \ref{figs11}(b) shows the square norm from Eq. 27 of the main text for different values of $\lambda$. The square norm for $\lambda = 0$ shows a spike at $r_e\approx 0.1$ and then increases gradually with aspect ratio as in Fig. \ref{figs11}(b). The square norm is of order $10^5$. This suggests values of $\lambda$ of order $10^{-1}$. The squares of the residual norm between the measured CLD $\mathbf{C}^{*}$ and calculated CLD $\mathbf{C}$ for $\lambda = [0.1, 0.2, 0.3]$ are shown in Fig. \ref{figs11}(a) while the corresponding squares of the norms of the recovered number based PSD are shown in Fig. \ref{figs11}(b).
 
 As expected, the spikes in the recovered number based PSDs are mitigated for $\lambda\neq 0$ as seen in Fig. \ref{figs11}(b). However, the penalty becomes less effective as the aspect ratio increases resulting in an increase in the squares of PSD norms in Fig. \ref{figs11}(b) with increasing $r_e$. Also, the fits to the experimental data reduces as $\lambda$ increases as seen in Fig. \ref{figs11}(a), but the mismatch increases with aspect ratio.
 
 The process of penalising spikes in the recovered number based PSD (at the cost of reduced match of the experimental data as seen in Fig. \ref{figs11}(a)) seen in Fig. \ref{figs11}(b) leads to the development of a global minimum in the objective function $f_2$ as seen in Fig. \ref{figs11}(c) for $\lambda\neq 0$. For $\lambda = 0.1$, the global minimum is quite shallow and not so obvious. However, it gets clearer at $\lambda = 0.2$ as in Fig. \ref{figs11}(c). The global minimum occurs at about the same region of $r_e\approx 0.3$ for $\lambda = [0.1, 0.2, 0.3]$ as in Fig. \ref{figs11}(c). Since $\lambda = 0.2$ yields a clear global minimum for the objective function $f_2$ with less cost on the quality of fit, then the value of $\lambda = 0.2$ was chosen for the minimisation of the objective function $f_2$ in the main text for the LW model. A similar procedure led to the choice of $\lambda = 0.01$ for the VSM model.
 
 The effect of penalising the number based PSD is shown in Fig. \ref{figs12}. At $r_e=0.1$ and $\lambda=0$, the recovered number based PSD has a long thin spike at $D_e\approx 1$ as in Fig. \ref{figs12}(a). However, the spike at $D_e\approx 1$ is removed for $\lambda\neq 0$ as seen in Fig. \ref{figs12}(a). The cost of removing the spike at $D_e\approx 1$ in Fig. \ref{figs12}(a) is the introduction of oscillations at small particle sizes. Similar to the spike at $D_e\approx 1$ in Fig. \ref{figs12}(a) is the spike at $D_e\approx 2$ in Fig. \ref{figs12}(b) (although shorter than the case of Fig. \ref{figs12}(a)) for $\lambda = 0$. This spike is removed for $\lambda\neq 0$. The same situation plays out at $r_e=0.3$ in Fig. \ref{figs12}(c). As the aspect ratio increases, the single long spike at a small particle size is replaced by small oscillations at small particle sizes for $\lambda = 0$ as seen in Figs. \ref{figs12}(c) to \ref{figs12}(f). However, at the optimum value of $\lambda = 0.2$ selected in Fig. \ref{figs11}, the distributions close to the minimum $r_e\approx [0.2,0.3]$ in Fig. \ref{figs11}(c) have the least oscillations at all particle sizes as seen in Figs. \ref{figs12}(b) and \ref{figs12}(c). This confirms that the choice of $\lambda = 0.2$ (for the LW model) used in Eq. 26 of the main text and the consequent minimum of the objective function $f_2$ at $r_e\approx 0.3$ for sample 1 in Fig. 11(c) yield physically realistic PSDs. A similar situation holds for the other samples and the VSM model. 
 
 Having obtained the optimum size ranges using the inversion algorithm and the optimum aspect ratio by minimising the objective function $f_2$, then the volume based PSD can be calculated at the optimum particle size range range and aspect ratio. The volume based PSD is calculated by minimising the objective function $f_3$ (defined in Eq. 33 of the main text) using the optimum particle size range and aspect ratio. The objective function $f_3$ could be minimised at $\lambda = 0$ or $\lambda\neq 0$ depending on the level of noisebin the recovered volume based PSD. For example, in the cases of samples 1 and 5 (using the LW model), the volume based PSDs recovered by minimising the objective function $f_3$ at $\lambda = 0$ contain spikes at $D_e\approx 200$ as in Fig. \ref{figs13}(a). This is because the corresponding number based PSDs contain small fluctuations at $D_e\approx 200$ leading to an exaggerated particle size counts at $D_e\approx 200$. These spikes at $D_e\approx 200$ in samples 1 and 5 can be removed by searching for a suitable value of $\lambda\neq 0$ so that the penalty on the norm of the PSD in the objective function $f_3$ becomes effective.   
  
    \begin{figure}[tbh]
     \centerline{\includegraphics[width=\textwidth]{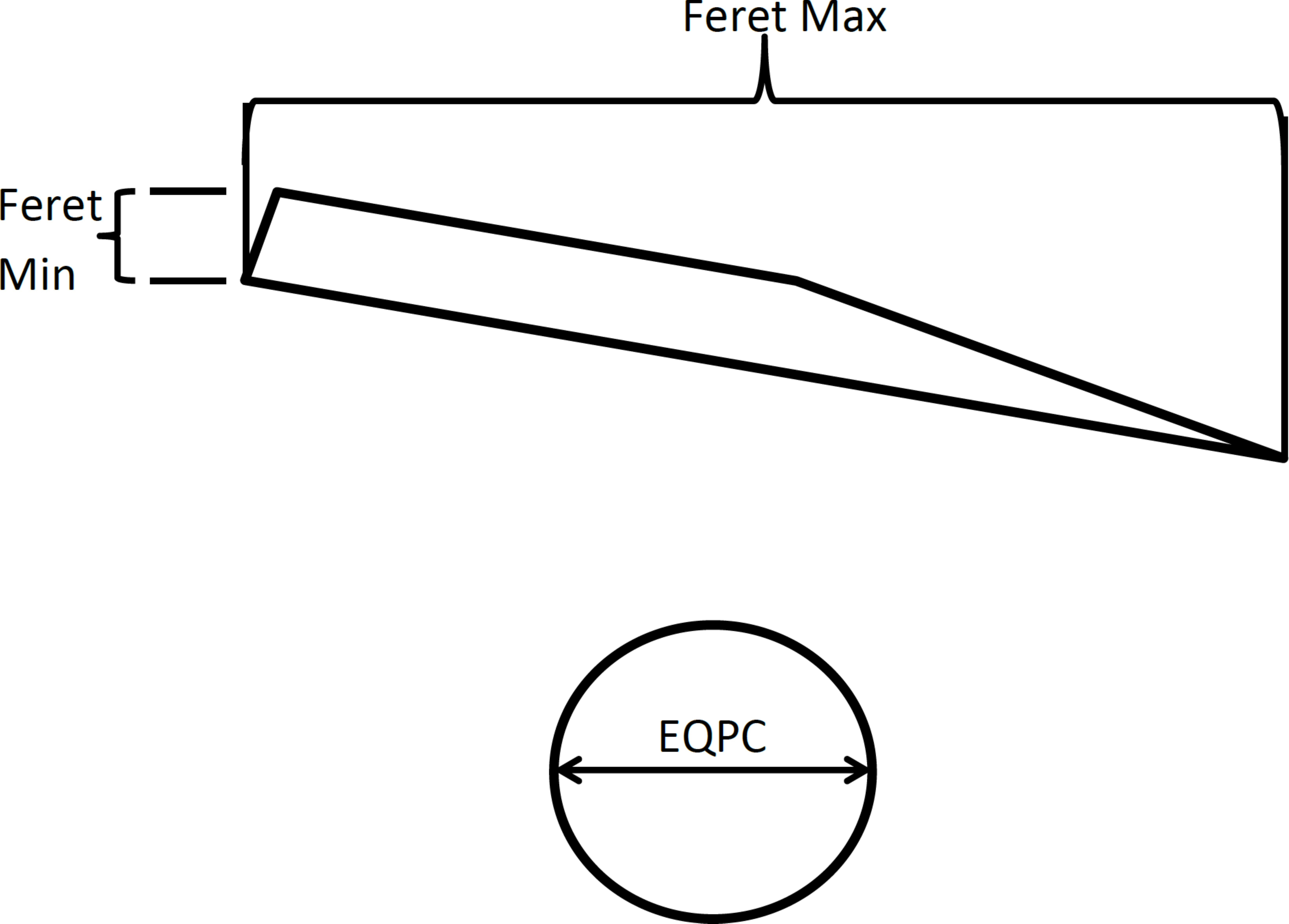}}
     \caption{Schematic representation of metrics Feret Max, Feret Min and EQPC obtained by dynamic image analysis for the five samples.}
      \label{figs14}
       \end{figure}
 
 The procedure for selecting $\lambda$ is similar to the case of the number based PSD. However, this time around the selection is done using the optimum particle size range obtained from the inversion algorithm and aspect ratio recovered from the objective function $f_2$. The sum of the squared deviation $[C_j^o-\overline{A}^o_{ji}\overline{X}^v_i]^2$ in Eq. 33 of the main text is of order $[10^{-6}, 1]$ for $r_e\in [0.15,0.4]$ for the five samples as seen in Fig. \ref{figs13}(d). The sum of the squares $[\overline{X}_i^v]^2$ in Eq. 33 is of order $[10^8, 10^{11}]$ for the five samples as in Fig. \ref{figs13}(c). This suggests values of $\lambda$ of order $10^{-15}$. The value of $\lambda = 8\times 10^{-15}$ was used for the five samples for calculations with the LW model. The resulting volume based PSDs for the five samples obtained at $\lambda = 8\times 10^{-15}$ are shown in Fig. \ref{figs13}(b). Figure \ref{figs13}(b) shows that the spikes at $D_e\approx 200\mu$m have been removed for samples 1 and 5.
  
 The volume based PSDs obtained by minimising the function $f_3$ (at $\lambda = 0$) with the VSM calculation 
 were fairly smooth unlike the case of Fig. \ref{figs13}(a). Hence the value of $\lambda=0$ was employed for the five samples for the case of the VSM model. The volume based PSDs recovered from the function $f_3$ (at $\lambda = 0$) using the VSM model for sample 1 to sample 5 are shown in Fig. \ref{figs13}(e).

 \section{Dynamic image analysis}
  
As mentioned in the main text, dynamic image analysis was performed with a QICPIC (Sympatec Ltd., UK) instrument with a LIXELL wet dispersion unit. The metrics obtained from dynamic image analysis were the equivalent projected circle EQPC diameter, the maximum feret (Feret Max) and the minimum feret (Feret Min) as described in \cite{Hamilton2012}. The Feret Max is the longest distance between two parallel tangents on opposite sides of the projected particle, while the Feret Min is the shortest distance between two tangents on opposite sides of the projected particle \cite{Hamilton2012}. The EQPC diameter is the diameter of a sphere whose 2 D projection has an area equal to the area of the 2 D projection of the particle.  These metrics are illustrated schematically in Fig. \ref{figs14}. 

\section{Possible discrepancies between calculated and measured chord lengths}

The calculation presented in the main text is based on a chord being defined as extending from edge to edge across a particle. The analytical models used in this work assume continuity of the particle boundary and the  geometry of each particle has been assumed to be perfectly cylindrical or ellipsoidal with no concavities. However, real convex (approximately) particles contain small concavities along their boundaries which implies that the particle boundaries are not always continuous or smooth. The approach used here implies that these small discontinuities in the particles' boundaries have been removed by using a linear interpolation between the points of discontinuity. 
 
 However, the presence of concavities along particles' boundaries will introduce small discrepancies between measured  and calculated chord lengths. The typical particle size is of the order of 100$\mu$m so that the depth of these concavities will be less than 1$\mu$m. Hence we expect an error of less than 1\% in the calculated chord lengths. However, for particles with more pronounced concavities (for example agglomerates) the error could increase significantly if the concavities are not properly accounted for by the model used. Work on a suitable model for dealing with agglomerates (which contain pronounced concavities) is currently in progress.
 
Another factor that can introduce discrepancies between the measured and calculated chord lengths is the optical properties of the particles. The focal spot of the laser has a fixed width. The laser beam converges towards the focal plane and diverges away from it. Hence a suitable threshold is used in the FBRM sensor to determine when the reflected light is accepted or rejected. This then implies that the length of a chord depends on the distance of the particle (from which light is reflected) from the focal plane and its reflecting properties. A small particle (whose size is close to the width of the laser spot) close to the focal plane could give rise to a measured chord length which is larger than its true value if the particle has very good reflectance. However, this small particle may be missed completely by the FBRM sensor if the particle is far away from the focal plane and has a poor reflectance. 
 The situation is similar for a large particle (whose size is significantly larger than the width of the laser spot). The measured chord length could be larger or smaller than the true value depending on the reflecting properties of the particle and its distance away from the focal plane. 
 
 Hence the optical properties of the particles in a population determine if a measured CLD is representative of the particles in the population or not. The degree of accuracy of the calculated CLD will also be affected by whether the optical properties of the particles are taken into account in the models or not. In the work presented here all particles are assumed to be opaque and to have good reflectance.  Representative images of the particles in Fig. 1 of the main text shows that this approximation is justified. Hence we do not expect a significant shift in the peak of the calculated CLD for this kind of system of particles. However, for a system of highly transparent particles, there could be a significant shift in the peak of the calculated CLD and hence the optical properties of the particles will need to be taken into account.

\end{document}